\documentclass[aps, prc, 12pt, nofootinbib, showpacs, superscriptaddress, tightenlines, groupedaddress]{revtex4-2}
\usepackage{amsmath,amssymb,amsbsy,bm}
\usepackage{graphicx}
\usepackage{comment}
\usepackage{float}
\usepackage[colorlinks=true,linkcolor=blue,citecolor=blue,urlcolor=blue]{hyperref}
\usepackage[margin=0.75in]{geometry}
\usepackage{silence}
\begin{document}

\title{Baryon Transport in Color Flux Tubes}
\author{Scott Pratt}
\affiliation{Department of Physics and Astronomy and Facility for Rare Isotope Beams\\
Michigan State University, East Lansing, MI 48824~~USA}
\date{\today}

\pacs{}

\begin{abstract}
Color flux tubes are a standard perspective from which to understand stopping in high-energy collisions. Mechanisms for baryon transport and polarization within a tube are considered here, both in regards to the transport of baryons from the target and projectile toward mid-rapidity, and in regards to the correlations of baryon-antibaryon pairs created in the tube. The roles of tube merging and gluon radiation with a tube are emphasized.
\end{abstract}

\maketitle

\section{Introduction}

Recent observations and ideas in proton-proton, proton-nucleus and nucleus collisions have renewed interest in baryon transport in hadronic collisions\cite{Brandenburg:2022hrp,STAR:2008med,STAR:2017sal,BRAHMS:2003wwg}. Baryons from the initial incoming nuclei tend to lose roughly one unit of rapidity in the collision. The distribution of baryons then appears to fall off exponentially as a function of rapidity loss, a result qualitatively consistent with the baryon-junction hypothesis, where baryon number is represented by gluonic degrees of freedom rather than solely by quarks \cite{Kharzeev:1996sq}. Naturally, this inspires one to re-examine less exotic pictures of baryon transport, which have the same qualitative behavior \cite{Vance:1997th}. The less exotic pictures, based on flux-tube or string models, have long been a staple of the stopping of baryons in hadronic collisions. The idea is that baryons, after exchanging a gluon, develop a tube of color electric flux that provides a force, or string tension acting on the baryon, thus slowing it down. As proposed in \cite{Vance:1997th}, when two strings, which originate from two separate quarks, merge into one, baryons are transported to the point of merging. The transport is driven by the polarization of quark-antiquark pairs which tunnel out of the vacuum in such a way that quarks consolidate to form a baryon at the point where the strings merge.

Here, the role of baryon transport in color fields or flux tubes is investigated within the context of the flux-tube picture. Much of this perspective is similar to the string picture outlined in \cite{Vance:1997th}. This approach is largely heuristic, but nonetheless demystifies the exponential fall-off of the stopping, and clarifies how a merging of flux tubes or strings is related to baryon transport. Even more sophisticated models of merging \cite{Sjostrand:2004pf}, or color recombination \cite{Christiansen:2015yqa}, are now applied in the PYTHIA 8 generator \cite{Bierlich:2022pfr}. The main idea is not completely orthogonal to that of the baryon junction paradigm outlined in \cite{Kharzeev:1996sq}, and the differences with the ideas presented here are more foundational than operational. For instance, for all the discussions in this paper, baryon number is manifestly carried by quarks. Here, the emphasis is on describing how color electric flux polarizes the baryon charge of the vacuum. Analogously, if one applies an electric field to a dielectric, the ensuing polarization leads to positive electric charge congregating on one side of the dielectric and negative charge on the opposite side. In classical electrodynamics one can describe the charge density as $\nabla\cdot\vec{E}$, but that does not mean that electric fields are carrying electric charge. Instead, dipoles have simply coordinated their alignment, with the act of charge separation briefly generating a current. Even though surface charge density appears on each side of the dielectric, no individual electric charges moved any distance further than the size of a dipole. The same can be done here, as even though baryon charge is ``moving'' along the tube, baryon number is still carried by quarks even though no quark or antiquark is required to travel any appreciable distance. Another feature of this picture, also qualitatively consistent with data, is that the transport, or stopping, of electric charge should be significantly weaker than that of baryons.

A central goal of this study is to better understand how color fields, which couple to color, not baryon number, induce baryon currents and baryon number polarization, and how the polarization results in baryon number traversing large swaths of rapidity. The next section considers the simplest version of a flux tube, one with a quark at one end and an antiquark on the other. This is mainly a restating of the ideas in \cite{Vance:1997th}. After reviewing how even in such a simple string the baryon number at the end of the strings vanishes into mesons, the merging of two simple strings into one is considered. It is shown that when two tubes from the two quarks in the projectile merge into one, with the new string connecting to a quark in the target hadron, that the baryon number from the projectile is transferred to the point of merging. Further, it is shown how this can be driven by simply exchanging a soft gluon between the target and projectile. Finally, it is pointed out that the most likely distribution of rapidity range of the baryon movement is exponential, assuming the paths are thermally weighted.

Section \ref{sec:kubo} is more formal, and describes how one can write a linear response theory connecting both baryon current and polarization to a field-like quantity based on standard QCD fields. The quantities that play the role of fields are shown to be gauge-invariant correlators involving both the gluonic field operators and the color charge currents driving the field. Charge and field operators related to the cubic Casimir of SU(3) are shown to play a pivotal role in baryon transport. Given that this physics is non-perturbative, linear response theory is questionable. Nonetheless, the exercise provides a basis for understanding how baryonic transport is driven by color fields. Even purely gluonic excitations are shown to have the ability to induce baryon currents.

A given point in a single flux tube can be characterized by the color multiplet defining the charge to each side of some given point. Section \ref{sec:compound} reviews how a tube characterized by a higher multiplet might decay to color singlets, and how such decays can induce baryon transport, perhaps of multiple baryons, across the tube. This sets the stage for Sec. \ref{sec:gluonradiation}, which extends the idea of color flux tubes to include intermediate color charges placed along the tube, from gluons. A simple model is presented where a flux tube is defined by the random color charges, from quarks at the ends and from gluons in between, with the constraint that the overall system is in a color singlet. The implications of such a picture in regards to baryon transport are discussed, along with the ramifications toward multiplicity distributions or correlations.

The potential observable consequences of these pictures are discussed in the summary. Observables are proposed that might test expectations of the paradigms presented here, but given the gross simplification and qualitative nature and the of the flux-tube paradigm, further study is needed before understanding whether the proposed measurements  might be sufficiently strong to warrant investing experimental effort. 

\section{Baryon Transport from Merging Simple Flux Tubes}\label{sec:simple}

\begin{figure}
\centerline{\includegraphics[width=0.6\textwidth]{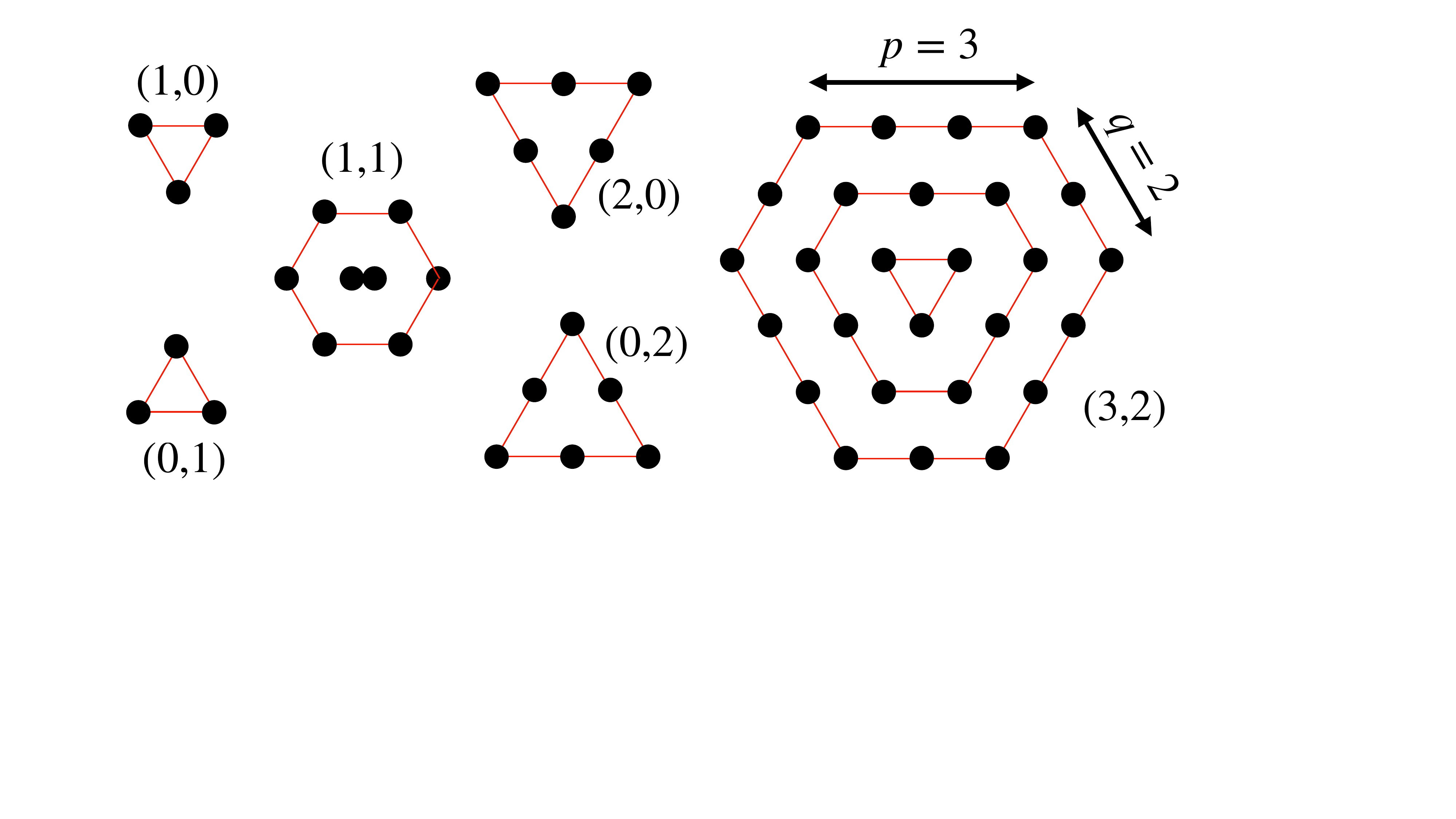}}
\caption{\label{fig:pqmultiplet}
Graphical representation of $(p,q)$ multiplets. The quark and antiquark color triplets are represented by $(1,0)$ and $(0,1)$, respectively, while the gluon octet is $(1,1)$. Each dot represents one projection of the multiplet, if the dot is on the outer ring. Subsequently, for the next inner-ring ring a dot represents two states, or three states for the next inner-ring, although the increasing degeneracy stops once a ring has either $p=0$ or $q=0$. The net degeneracy of the multiplet is $(p+1)(q+1)(p+q+2)/2$. In SU(2) one can combine multiplets of $j_1$ and $j_2$ to form several multiplets $j'$. Similarly, in SU(3) one can combine multiplets of $(p_1,q_1)$ and $(p_2,q_2)$ to form multiplets of several combinations of $(p',q')$.
}
\end{figure}

Here, we consider the merging of two simple flux tubes into one, similarly to the picture in \cite{Vance:1997th}. The term ``simple'' refers to the fact that the tube is generated by a single quark or anti-quark at the end opposite the merging point. In a simple tube, one can bisect the tube at any point and classify the color charge on each side of the dividing point by its color multiplet, $(p,q)$. In SU(3), the multiplet is defined by two integers, which differs from SU(2), where a multiplet is denoted by a single number, $j$, i.e. the total angular momentum. Graphically, the state can be classified by the graphs in Fig. \ref{fig:pqmultiplet}. Multiplets denoted by $(p_1,q_1)$ and $(p_2,q_2)$ can be combined to create multiplets denoted by $(p',q')$. In SU(2), multiplets denoted by $j_1$ and $j_2$ only couple to a singlet, $j'=0$, if $j_1=j_2$. Similarly, for SU(3) two multiplets couple to a color singlet only if $p_1=q_2$ and $p_2=q_1$. For example, the $(1,0)$ multiplet (quark) and the $(0,1)$ multiplet (antiquark) can couple to either a color singlet or to an octet, $(p'=1,q'=1)$.

\begin{figure}
\centerline{\includegraphics[width=0.4\textwidth]{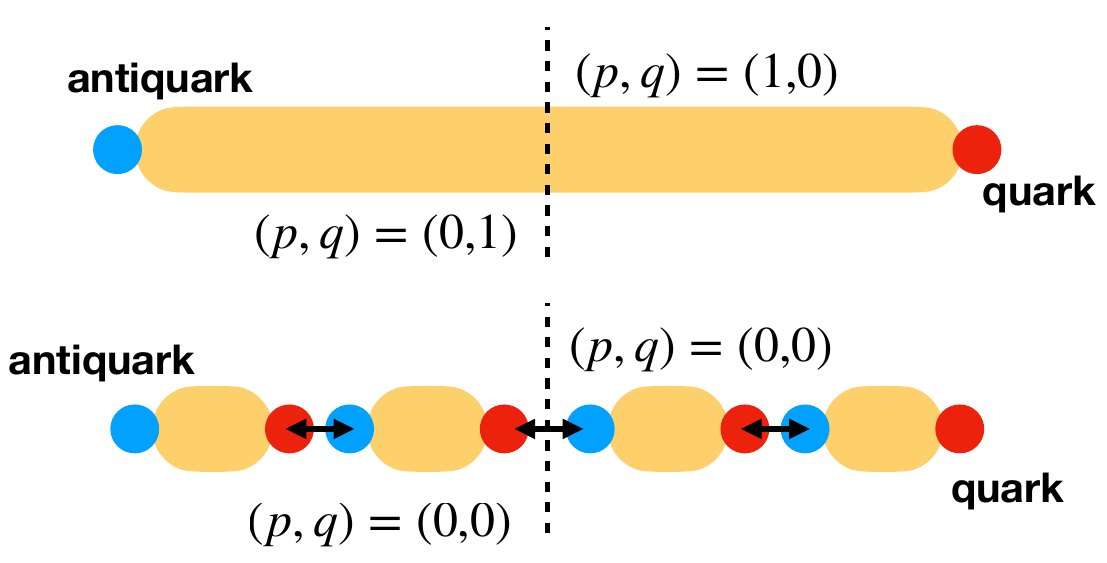}}
\caption{\label{fig:simpletube}
A simple flux tube between a quark $(p=1,q=0)$ and an antiquark $(p=0,q=1)$ is illustrated in the upper panel, with the lighter and darker circles representing the quark and antiquark respectively. Drawing a vertical line through a flux tube divides space into two regions, the left side of the dashed line in a color multiplet $(p=1,q=0)$ and the right side in the multiplet $(p=0,q=1)$. The lower panel illustrates how quark-antiquark pairs can tunnel out of the vacuum so that the space between the tunneling quarks is free of color fields. Tunneling quark-antiquark pairs are denoted by the arrows.  Dividing space in such a way that the dividing line does not bisect a tube results in both sides being in color singlets.
}
\end{figure}

Before addressing merging, we review the case of a simple flux tube between a quark and antiquark. Figure \ref{fig:simpletube} illustrates how the flux can be reduced by quark-antiquark pairs tunneling out of the vacuum. One can pick any position along the tube (denoted by a dashed line in the figure), and after the tube has decayed via quark-antiquark pair production, all matter to either side of the position should return to a color singlet. This is most easily accomplished by creating a $q\bar{q}$ pair out of the vacuum, with the antiquark moving toward the side of the tube that had initially had a single quark, and the quark moving to the opposite side. To create a color singlet each side of the designated position, either an antiquark must cross the dashed line, moving from right to left, or a quark must cross the  line moving the opposite direction. Creating the pair requires energy, but by separating sufficient distance the reduced field energy between the quark-antiquark pair can more than account for the masses and kinetic energies of the tunneled quarks. If the string tension, and energy per unit length, is $\lambda$, the field energy gained by producing the pair is $\lambda \ell$, where $\ell$ is the distance separating the quark and the antiquark. This process is sometimes compared to Schwinger pair production \cite{Schwinger}, where $e^+e^-$ pairs can be created in sufficiently strong electric fields. The QCD process differs in that the flux tube energy, $\lambda\ell$, assists with the pair production, whereas in the constant electric field case, one only has $eE\ell$, where the electric field is $E$. The Schwinger effect has been proposed to exist in regions of extremely high electric field in plasmas \cite{SchwingerPlasma}, in graphene \cite{SchwingerGraphene}, where the electrons and holes become nearly massless, as well as in the context of hadronic collisions \cite{Wong:1994ei,Suganuma:1991ha}.

A simple flux tube, with a quark at one end and an antiquark on the other, typically decays completely into mesons. Even though each end of the string has baryon number, and even though there are no baryons in the final state, no quark moved any significant distance along the tube. Instead, quark-antiquark dipoles appeared, effectively shifting a line of quarks in one direction and a line of antiquarks in the opposite direction. The act of shifting provides a current ephemerally. This is similar to placing a dielectric in an electric field. Opposite charge densities appear on each side of the dielectric even though no individual charge moved further than the size of a dipole. The baryon current involved can be equated to the time rate of change of the polarization. One way in which a color flux tube differs from electric field is that the induced color current will persist until the field is completely neutralized, unlike in a dielectric, where the current persists until the induced polarization reduces the initial field by the permeability of the matter.

\begin{figure}
\centerline{\includegraphics[width=0.48\textwidth]{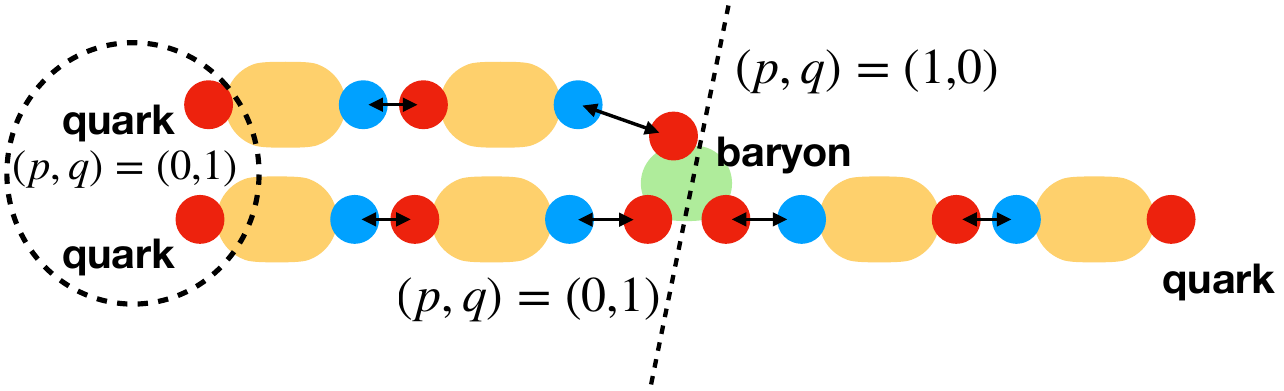}}
\centerline{\includegraphics[width=0.6\textwidth]{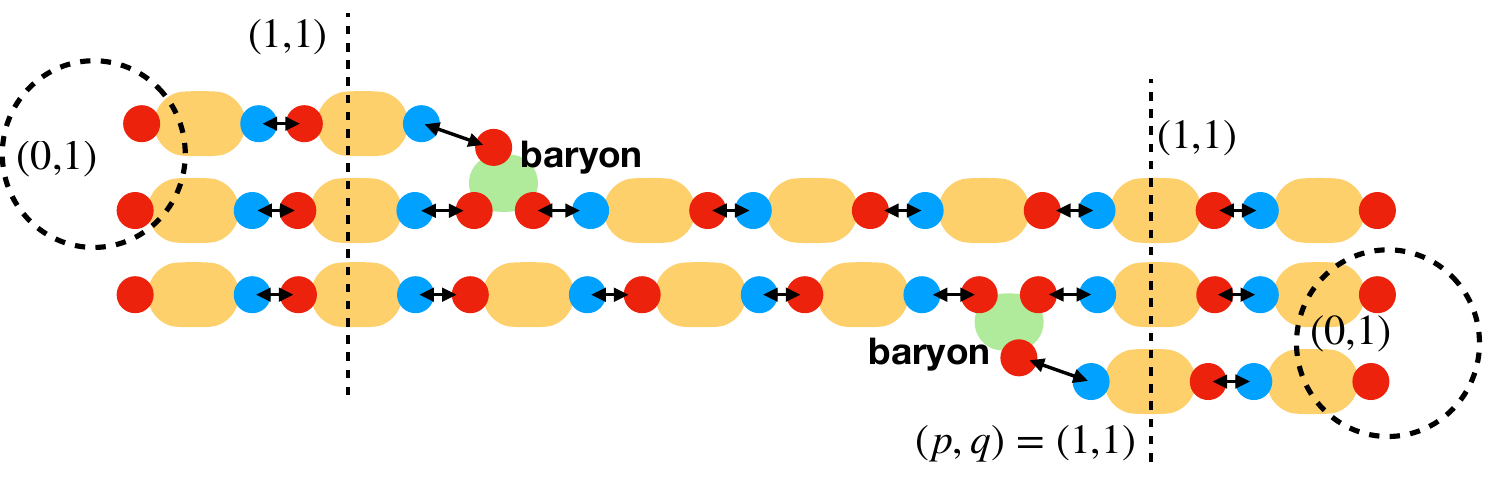}}
\caption{\label{fig:merge}
Upper illustration: The merging of two tubes into one, where the left-hand side tubes begin with two quarks, and the right-hand side ends with one quark. The two quark multiplets couple to a $(0,1)$ multiplet, which then matches with the quark multiplet on the right. The baryon number, is then located at the point where the two tubes merged. This represents a situation where one quark is scattered far, from its original momentum, and the flux tubes must adjust to form local color singlets. The original quark pair must couple to $(p=0,q=1)$ if the overall configuration is to be a color singlet.\\
Lower illustration: This represents the case where two baryons exchange a soft gluon. In that case both of the baryons are transformed into $(p=1,q=1)$ multiplets. Assuming that for each baryon there is a diquark with $(p=0,q=1)$, the overall color singlet is restored by migrating the baryon number to the two places where the strings merge.\\
Red (or lighter gray) circles represent quarks, while blue (or darker gray) circles represent antiquarks. Ovals represent fluxtubes, and arrows describe quark-antiquark pairs that tunneled from vacuum.
}
\end{figure}
Figure \ref{fig:merge} shows how baryon number can move from the target or projectile rapidity toward central rapidity through merging of simple flux tubes. As was the case for a single tube with a quark at the end, quark-antiquark pairs appear in a polarized manner. The baryon number of 1/3 at the end of the string then gets absorbed into mesons. Following the string, once the mesons appear, there is a quark left over on the string. This leftover quark represents the baryon transport. If three strings merge, the three remaining quarks can produce a baryon. However, if the three quarks originated from the same baryon, e.g. a projectile proton, and if that proton exchanged a gluon with the target, then those three quarks would then be in a color octet, $(p=1,q=1)$, and they could not form a baryon. However, if one of the three merging tubes came from the opposite side, then a baryon can form at the merging point, as illustrated in the upper diagram of Fig. \ref{fig:merge}.

The lower diagram of Fig. \ref{fig:merge} illustrates how all six quarks can recombine through string merging. Thus, baryon number should move inward from both directions. Of course, the most energetically favorable path is for the tubes to merge as quickly as possible. When two quarks merge into a single tube, and if the two quarks were coupled to a $(p=0,q=1)$ state, that is known as a diquark. Once merged the tube should have the same field energy as the tube generated by a single quark or antiquark. If the system chooses the point to merge based on local thermal penalties, one would expect the probability for merging to fall exponentially with the separation from the initial rapidity of the baryon. Phenomenologically, one expects approximately one tube breaking per unit of rapidity. If the characteristic energy of the string is similar to the characteristic temperature, one might expect the characteristic exponential falloff to be of order unity, but it should be emphasized that it would not be surprising if the actual exponential scale was a half unit of rapidity. The chance that a baryon might be transported from the target or projectile rapidity, a difference of 7 units at the LHC, would be extremely sensitive to this scale.

Figure \ref{fig:merge} illustrates two tubes merging that began at the target or projectile nucleons. One might also consider such tubes splitting again, as one moves further towards mid-rapidity, perhaps splitting and merging more than once. In that case if a tube splits and merges, a baryon would be at one end and an antibaryon at the other. There would be no preference for baryon to appear before or after the anti-baryon. If this were the mechanism for how baryon-antibaryon pairs are created from flux tubes, one would expect that the balancing pair's separation in rapidity to fall off exponentially, with the same exponential falloff as that characterizing the transport of baryon number from the target or projectile. Observing such separation would require experimental particle identification across multiple units of rapidity.

In addition to providing additional baryon number, if the projectile and target are both protons, they also provide additional electric charge. In a flux tube the dipole-like tunneling quarks line up with the antiquark on the side of the quark at which the tube originated. However, for electric charge, the tunneling pair would roughly have equal probability of being a $u\bar{u},~d\bar{d}$ or $s\bar{s}$ pair. This would result in zero polarization of the electric charge. If the $s\bar{s}$ pairs were slightly less, then a small electric charge transport would ensue. It would be smaller in magnitude than that for baryon number, but would be characterized by the same exponential falloff. 

\section{Connecting Baryon Transport to Color Fields}\label{sec:kubo}

It is not obvious to see how baryon transport is connected to gluonic fields, $G^{\mu\nu}_a(x)$. Unlike the case for electric fields, gluon fields have a color index, $a$, and because all of nature is in a color singlet, the expectation of the field operators are zero, $\langle G^{\mu\nu}_a(x)\rangle=0$. Therefore, one must describe the fields through operators involving products of operators with color indices, i.e. corrrelators involving two or more operators with color indices. Further, gluons couple to color charge, not to baryon charge. The rather modest goal of this section is to write operators in terms of QCD field operators that represent the baryon polarization or baryon current and the field-like operators that drive the polarization or current. These operators will be discussed in the context of flux tubes. None of the expressions derived here will be applied to quantitatively predict observables in this study. However, the presentation may help elucidate how fields in QCD can inspire baryon polarization or transport.

\subsection{Color-Neutral Correlators}

Here, the color electric/magnetic field operators and the color current operators will be considered, $G^{\mu\nu}_a(x)$, and $j^\mu_a(x)$. Instead of the usual definition of such operators, we add an additional feature that enables the construction of gauge-invariant correlators. For any operator $\mathcal{O}_a(x)$, an additional attached operator is inferred \cite{Elze:1989un},
\begin{eqnarray}
\mathcal{O}_a(x)&\rightarrow \left[\mathcal{P}\exp\{ig\int d\vec{\ell}\cdot\vec{A}\}\right]\mathcal{O}_a(x)\left[\mathcal{P}\exp\{-ig\int d\vec{\ell}\cdot\vec{A}\}\right].
\end{eqnarray}
Here, $\mathcal{P}$ is the path-ordering operator. With this definition, if some correlator $\langle\mathcal{O}_a(x)\mathcal{O}_a(x)\rangle$ is gauge invariant, then $\langle\mathcal{O}_a(x)\mathcal{O}_a(y)\rangle$ is also gauge invariant, though one needs to realize that the new correlator might depend on the actual path chosen unless the path integrals on each side of the operator return via the same path. With this definition, one can write expressions that appear similar to those for Maxwell's equations,
\begin{eqnarray}
\partial_\mu j^\mu_a(x)&=&0,\\
\nonumber
\partial_\mu G^{\mu\nu}_a(x)&=&j^\nu_a(x).
\end{eqnarray}
Because all observables must be invariant to rotations in color space, one knows that $\langle\mathcal{O}_a\rangle=0$ and one must consider only colorless correlators. To find quantities which can be treated, at least to some degree, similarly as fields, we consider the following operators defined in terms of operators $j_a^\mu(x)$, where $a$ is a color index. 
\begin{eqnarray}
Q_{a,\Omega}&=&\int d\Omega_\mu j^{\mu}_a(x),\\
\nonumber
\langle Q^{(2)}_{\Omega_1,\Omega_2}\rangle&=&\int d\Omega_{1,\mu}d\Omega_{2,\nu}\langle j^\mu_a(x_1) j^\nu_a(x_2)\rangle,\\
\nonumber
&=&\langle Q_{a,\Omega_1}Q_{a,\Omega_2}\rangle,\\
\nonumber
\langle j^{(2)\nu}_{\Omega_1}(x)\rangle &=&\int d\Omega_{1,\mu} \langle j^\mu_a(x_1) j^\nu_a(x)\rangle\\
\nonumber
&=&\langle Q_{a,\Omega_1}j^\nu_a(x)\rangle.
\end{eqnarray}
Here, the superscripts $(2)$ reference the fact that these correlations are quadratic in the color charge. The volumes $\Omega_1$ and $\Omega_2$ can be any chosen volumes, and might be the same volume. If the volumes cover all space, and if the system is in an overall color singlet, $\langle Q^{(2)}\rangle=0$. One can also define correlators using the fields,
\begin{eqnarray}
\langle G^{(2)\mu\nu}_{\Omega}(x)\rangle&=&\int d\Omega\langle Q_{a,\Omega}G^{\mu\nu}_a(x)\rangle.
\end{eqnarray}

As mentioned above, the correlators $\langle\mathcal{O}_a(x)\mathcal{P}_a(x)\rangle$ are gauge-invariant for any operators $\mathcal{O}$ or $\mathcal{P}$ that transform as the color charge, or equivalently as the 8 generators in SU(3). In fact, there are other combinations of operators that transform as color singlets. SU(3) has two Casimir operators, a Casimir being some combination of the group generators that commutes with the eight generators, $\lambda_a$. The first Casimir, often referred to as the quadratic Casimir
\begin{eqnarray}
C^{\rm(quadratic)}&=&\sum_a\lambda_a\lambda_a,
\end{eqnarray}
corresponds to the two-body correlators above. Unlike SU(2), SU(3) has a second Casimir, the cubic Casimir,
\begin{eqnarray}
C^{\rm(cubic)}&=&\sum_{abc}d_{abc}\lambda_a\lambda_b\lambda_c,
\end{eqnarray}
where $d_{abc}$ are the symmetric structure constants.
\begin{eqnarray}
\{\lambda_a,\lambda_b\}&=&\frac{4}{3}\delta_{ab}+d_{abc}\lambda_c.
\end{eqnarray}
It is this second Casimir that will be associated with baryon transport. One can define a second set of current and charge-like operators, based on the cubic Casimir,
\begin{eqnarray}
\langle j^{(3)\mu}_{\Omega_1,\Omega_2}(x)\rangle &=&d_{abc}\langle Q_{a,\Omega_1}Q_{b,\Omega_2}j^\mu_c(x)\rangle,\\
\nonumber
\langle Q^{(3)}_{\Omega_1,\Omega_2,\Omega_3}\rangle&=&d_{abc}\int d\Omega_{1,\mu}d\Omega_{2,\nu}d\Omega_{3,\eta}\langle j^\mu_a(x_1) j^\nu_b(x_2)j^\eta_c(x_3)\rangle\\
\nonumber
&=&d_{abc}\langle Q_{a,\Omega_1}Q_{b,\Omega_2},Q_{c,\Omega_3}\rangle.
\end{eqnarray}
Again, the three volumes might or might not be the same. One can define the corresponding correlations involving field operators,
\begin{eqnarray}
\langle G^{(3)\mu\nu}_{\Omega_1,\Omega_2}(x)\rangle&=&d_{abc}\langle Q_{a,\Omega_1}Q_{b,\Omega_2}
G^{\mu\nu}_c(x)\rangle.
\end{eqnarray}

For the purposes of this discussion, two vector quantities are defined,
\begin{eqnarray}
E^{(2)}_{\Omega,i}(x)&=&\langle G^{(2)0i}_{\Omega}(x)\rangle,\\
\nonumber
E^{(3)}_{\Omega_1,\Omega_2,i}(x)&=&\langle G^{(3)0i}_{\Omega_1,\Omega_2}(x)\rangle.
\end{eqnarray}
These represent two measures of the color electric field, but are in fact correlators. They have no color index, as their color is defined relative to the color of the matter within the volume $\Omega$. These spatial vectors participate in two versions of Gauss's Law,
\begin{eqnarray}\label{eq:gauss}
 \oint d\vec{S}_2\cdot\vec{E}^{(2)}_{\Omega_1}(x_2)&=&Q^{(2)}_{\Omega_1,\Omega_2},\\
 \nonumber
 \oint d\vec{S}_3\cdot\vec{E}^{(3)}_{\Omega_1,\Omega_2}(x_3)&=&Q^{(3)}_{\Omega_1,\Omega_2,\Omega_3}.
\end{eqnarray}
Again, the volumes, $\Omega_1,\Omega_2,\Omega_3$, might or might not be chosen as being the same.

For a flux tube, where the fields are constant across a cross section $A$, one can define a volume $\Omega$ as all tube to the left of some position where we bisect the tube. If the charge within that volume is described by $Q^{(2)}_{\Omega,\Omega}$, Gauss's law states that the field is 
\begin{eqnarray}
E^{(2)}_{\Omega,\Omega}&=&\frac{1}{A}Q^{(2)}_{\Omega,\Omega}\\
\nonumber
&=&\langle Q_{a,\Omega}E_a\rangle.
\end{eqnarray}
For a single charge $Q_a$, $\langle Q_{a,\Omega}Q_{a,\Omega}\rangle=\langle Q^{(2)}_{\Omega,\Omega}\rangle$. One can relate the energy density at some point in the flux tube to the fields,
\begin{eqnarray}
 \epsilon&=&\frac{1}{2}\langle E_a^2\rangle\\
 \nonumber
 &=&\frac{1}{2}\frac{(E^{(2)}_{\Omega,\Omega})^2}{Q^{(2)}_{\Omega,\Omega}}.
\end{eqnarray}
The energy per unit length, or string tension is then
\begin{eqnarray}\label{eq:tension}
\frac{E}{L}&=&\frac{A}{2}\frac{(E^{(2)}_{\Omega,\Omega})^2}{Q^{(2)}_{\Omega,\Omega}}\\
\nonumber
&=&\frac{1}{2A}Q^{(2)}_{\Omega,\Omega}.
\end{eqnarray}
Thus, $Q^{(2)}$ is a measure of the energy per unit length of the tube, although if the area changes with the strength of the field, the correspondence is not purely linear.

At the end of this section it will be shown how $Q^{(3)}$ is related to baryon number. The meaning of the two charges are, roughly, that $Q^{(2)}$ describes the energy per unit length of the flux tube and that $Q^{(3)}$ represents the propensity to attract baryons vs. antibaryons.

\subsection{Kubo Relations for Conductivity and Polarizability}

Here, we consider a system divided into two volumes, $\Omega$ and a small adjacent volume $\delta\Omega$. We will consider the currents and polarizability in $\delta\Omega$, while using $\Omega$ to define the correlators. The field, $\vec{E}^{(3)}(\vec{r}\in\delta\Omega)$ and the current $\vec{j}^{(3)}(\vec{r}\in\delta\Omega)$ will be defined by the charge distribution in $\Omega$. First, to derive the Kubo relation for the conductivity, we consider the interaction with a color field,
\begin{eqnarray}
H^{\rm(int)}&=&-\int d^3r \rho_a(\vec{r})\vec{r}\cdot\vec{E}_a.
\end{eqnarray}
Following the usual steps for deriving the first-order perturbation correction to the current,
\begin{eqnarray}
\langle\vec{j}^{(3)}_{\Omega\Omega}(\vec{r})\rangle&=&-i\int d^3r'\int_{-\infty}^0 dt~\langle[\vec{j}^{(3)}(\vec{r}),
\int d^3r'~\rho_a(\vec{r}',t)\vec{r}'\cdot\vec{E}_a(\vec{r}',t)]\rangle\\
\nonumber
&=&id_{abc}\int d^3r'\int_{-\infty}^0 dt'~\langle[Q_{a,\Omega}Q_{b,\Omega}\vec{j}_c(\vec{r}),
\int d^3r'~\rho_d(\vec{r}',t')\vec{r}'\cdot\vec{E}_d(\vec{r}',t')]\rangle.
\end{eqnarray}
Here, $\vec{r}$ is inside $\delta\Omega$. Whereas the usual step in deriving Kubo relations is to treat the field as a constant external applied field, here the operator $\vec{E}_{\Omega,\Omega}^{(3)}(\vec{r}')$ is factored out instead. Only the $c=d$ term will contribute in this way.
\begin{eqnarray}
\langle\vec{j}^{(3)}_{\Omega\Omega}(\vec{r})\rangle&=&i\int_{-\infty}^0dt'~\int d^3r'\vec{r}'\cdot
\langle[\vec{j}_a(\vec{r}),\rho_a(\vec{r}',t')]\rangle  \vec{r}'\cdot\vec{E}^{(3)}_{\Omega,\Omega}({\rm in}~\delta\Omega).
\end{eqnarray}
Aside from the color indices, the prefactor to the electric field is the same as for calculating the Kubo relation for a normal electromagnetic field. In terms of anti-commutators, the result looks like the familiar result for electric conductivity,
\begin{eqnarray}\label{eq:kubo}
\langle\vec{j}^{(3)}_{\Omega\Omega}(\vec{r})\rangle&=&\sigma \vec{E}^{(3)}_{\Omega,\Omega}(\vec{r}),\\
\nonumber
\sigma(\vec{r})&=&\frac{1}{2T}\int_{-\infty}^\infty dt'\int d^3r'~\langle\{\vec{j}_{ax}(\vec{r},t=0),\vec{j}_{ax}(\vec{r}',t')\}\rangle.
\end{eqnarray}
Thus, the conductivity, $\sigma$, doesn't include any mention of $\Omega$, and only depends on the local properties.

It should be emphasized that Eq. (\ref{eq:kubo}) describes the contribution to the current due to the charge distribution in $\Omega$. If $\Omega$ is a subvolume, then other regions not only contribute, but their charges might interfere, either constructively or destructively, with those in $\Omega$.

If quarks and gluons are free, one can have currents and fields as presented above. However, if the world is comprised of color singlets, there is no current, but there can be polarization. As will be discussed in the next subsection, the current $\vec{j}^{(3)}$ is correlated to a baryon current. Similarly, one can define a polarization of the color fields, that is also correlated to the polarization of baryon charge. Here, operators that play the role of polarization, and are also based on the cubic Casimir, are presented. 

The polarization operator is defined here as
\begin{eqnarray}
\vec{\mathcal{P}}_{\Omega_1,\Omega_2}^{(3)}(\vec{r})=\vec{r}d_{abc}\langle Q_{a,\Omega_1}Q_{b,\Omega_2}\rho_c(\vec{r})\rangle.
\end{eqnarray}
Using the same interaction as was used for the calculation of the conductivity above, one can calculate the thermal expectation of the polarization. Setting $\Omega_1=\Omega_2=\Omega$, and factoring out $E^{(3)}$ as was done earlier, one finds
\begin{eqnarray}\label{eq:polarization}
\langle\vec{\mathcal{P}}_{\Omega,\Omega}^{(3)}(\vec{r})\rangle&=&\kappa \langle\vec{E}^{(3)}_{\Omega,\Omega}(\vec{r})\rangle,\\
\nonumber
\kappa(\vec{r})&=&\frac{1}{3T}\int d^3r'\langle \rho_a(\vec{r})[\vec{r}\cdot\vec{r}']\rho_a(\vec{r}')\rangle.
\end{eqnarray}
As was the case for the conductivity, this expression for the polarizability does not depend on the volume $\Omega$. If the charges appear in uncorrelated $\pm$ pairs, the polarizability becomes $g^2\langle(x_+-x_-)^2\rangle\rho_d$, where $\rho_d$ is the density of dipoles and $x_+-x_-$ is the distance between the two charges, $\pm g$. 

\subsection{Relation to Baryon Transport}

As was promised previously, the goal of this subsection is to demonstrate how baryon transport is related to the field $\vec{E}^{(3)}$, which was in turn related to the cubic Casimir. The quadratic and cubic Casimirs can be expressed in terms of the color multiplet labels $p$ and $q$ through
\begin{eqnarray}\label{eq:Q2Q3vspq}
C^{\rm quadratic}&=&(p^2+q^2+3p+3q+pq)/3,\\
\nonumber
C^{\rm cubic}&=&(p-q)(2p+q+3)(2q+p+3)/18.
\end{eqnarray}
The correlators $Q^{(2)}$ and $Q^{(2)}$ are given by the Casimirs,
\begin{eqnarray}
\langle p,q|Q^{(2)}_{\Omega,\Omega}|p,q\rangle&=&g^2C^{\rm quadratic},\\
\nonumber
\langle p,q|Q^{(3)}_{\Omega,\Omega,\Omega}|p,q\rangle&=&g^3C^{\rm cubic},
\end{eqnarray}
where $g$ is the coupling constant, which plays the role of the fundamental charge, the analog of $e$ in electromagnetism. For a quark, $(p=1,q=0)$, one finds that $\langle Q^{(2)}\rangle=4/3$ and $\langle Q^{(3)}\rangle=(10/9)g^3$. An antiquark yields the opposite expectation,  $\langle Q^{(3)}\rangle=-(10/9)g^3$.

The multiplet labels $p$ and $q$ and the charge labels $Q^{(2)}$ and $Q^{(3)}$ represent alternative means to identify the multiplet. The labels $p$ and $q$ can be uniquely expressed in terms of the charges by solving a cubic equation. Defining
\begin{eqnarray}
c&\equiv&9\langle Q^{(2)}/g^2\rangle+9,\\
\nonumber
d&\equiv&18\langle Q^{(3)}/g^3\rangle,\\
\nonumber
\alpha&=&-2\pi /3+\frac{1}{3}\cos^{-1}[(3d/2c)\sqrt{3/c}],\\
\nonumber
y&=&2\sqrt{c/3}\cos\alpha,\\
\nonumber
x&=&-2+\sqrt{4+\langle Q^{(2)}/g^2\rangle-y^2/3},\\
\nonumber
p&=&(x+y)/2,\\
\nonumber
q&=&(x-y)/2.
\end{eqnarray}
The other two solutions to the cubic equation result in either $p$ or $q$ being negative. The solutions are unique in that no two combinations of  $Q^{(2)}$ and $Q^{(3)}$ result in the same $p$ and $q$.

Given that a quark has $p=1,q=0$ and an antiquark has $p=0,q=1$, it is natural to expect that a multiplet with $p>q$ will more likely dissolve into quarks, and that a multiplet with $q>p$ would more likely result in antiquarks. Given that $Q^{(3)}$ is an odd function of $p-q$ and that $Q^{(2)}$ is an even function, any correlation with baryon number is through the correlation with $Q^{(3)}$. Correspondingly, the field $E^{(2)}$ does not drive baryon transport, but $E^{(3)}$ does.

To express baryon transport, one must first understand $d\langle B\rangle/d\langle Q^{(3)}\rangle$. Once this is determined, the baryon current is
\begin{eqnarray}
\vec{j}_B(\vec{r})&=&\frac{d\langle B_{\delta\Omega}\rangle}{dQ^{(3)}_{\Omega,\Omega,\delta\Omega}}\sigma \vec{E}^{(3)}_{\Omega,\Omega}(\vec{r}).
\end{eqnarray}
Here, $\vec{r}\in\delta\Omega$. This represents the current inside $\delta\Omega$ driven by the charge in $\Omega$. Statistical considerations give
\begin{eqnarray}
\frac{d\langle B_{\delta\Omega}\rangle}{dQ^{(3)}_{\Omega,\Omega,\delta\Omega}}&=&
\frac{\langle B_{\delta\Omega} Q^{(3)}_{\Omega,\Omega,\delta\Omega}\rangle}
{\langle [Q^{(3)}_{\Omega,\Omega,\delta\Omega}]^2  \rangle}.
\end{eqnarray}
One can follow the same arguments to calculate the baryon polarizability operator,
\begin{eqnarray}
\vec{P}_B(\vec{r})&=&\vec{r}\langle \rho_B(\vec{r})\rangle.
\end{eqnarray}
The ratio of $P_B$ to $P^{(3)}$ is the same as would be used for the current above. The baryon current and polarizability are then
\begin{eqnarray}
\langle\vec{j}_B(\vec{r})\rangle&=&\frac{d\langle B_{\delta\Omega}\rangle}{dQ^{(3)}_{\Omega,\Omega,\delta\Omega}}\sigma
\langle\vec{E}^{(3)}_{\Omega,\Omega}(\vec{r})\rangle,\\
\nonumber
\langle\vec{P}_B(\vec{r})\rangle&=&\frac{d\langle B_{\delta\Omega}\rangle}{dQ^{(3)}_{\Omega,\Omega,\delta\Omega}}\kappa
\langle\vec{E}^{(3)}_{\Omega,\Omega}(\vec{r})\rangle.
\end{eqnarray}

\subsection{Scaled Correlators}

The operators $Q^{(2)}, Q^{(3)},E^{(2)}(\vec{r})$ and $E^{(3)}(\vec{r})$ are correlators. If one assigns a dimension to the charge coupling $g$, the units are $g^2$, $g^3$, $g^2/L^2$ and $g^3/L^2$, where $L$ refers to units of length. The operator $Q^{(2)}_{\Omega,\Omega}$ represents a fluctuation, and if one has a bulk system with independent color charges, and no long-range correlation, it would scale with the volume $\Omega$. If one were to consider $Q^{(3)}_{\Omega,\Omega,\delta\Omega}$ and the two field-like operators $E^{(2)}_{\Omega}(\vec{r})$ and $E^{(3)}_{\Omega,\Omega}(\vec{r})$, one would see that for independent charges, the fluctuations of these operators would scale, in the large $\Omega$ limit, as
\begin{eqnarray}
\langle [E^{(2)}_{\Omega}(\vec{r})]^2\rangle\propto \Omega,\\
\nonumber
\langle [Q^{(3)}_{\Omega,\Omega,\delta\Omega}]^2\rangle \propto \Omega^2\delta\Omega,\\
\nonumber
\langle [E^{(3)}_{\Omega,\Omega}(\vec{r})]^2\rangle\propto \Omega^2.
\end{eqnarray}
The correlator with baryon number would scale as
\begin{eqnarray}
\langle Q^{(3)}_{\Omega,\Omega,\delta\Omega}B_{\delta\Omega}\rangle\propto \Omega\delta\Omega.
\end{eqnarray}
Thus, if one were to consider a long flux tube, with $\Omega$ defined as everything to one side, the fluctuation of the field operators at the center of the tube would increase with the size of the tube, and the ratio of $d\langle B\rangle/d\langle Q^{(3)}\rangle$ would fall with size of the tube. 

One can scale the three operators above so that their dimensions and scaling with size are in line with the charge and field operators with which we are more accustomed.
\begin{eqnarray}
\tilde{\vec{E}}^{(2)}_{\Omega}(\vec{r})&\equiv&\frac{\vec{E}^{(2)}_{\Omega}(\vec{r})}{\langle Q^{(2)}_{\Omega,\Omega}\rangle^{1/2}},\\
\nonumber
\tilde{\rho}^{(3)}_{\Omega,\Omega}(\vec{r})&\equiv&\frac{\rho^{(3)}_{\Omega,\Omega}(\vec{r})}{\langle Q^{(2)}_{\Omega,\Omega}\rangle},\\
\nonumber
\vec{\tilde{\mathcal{P}}}_{\Omega,\Omega}^{(3)}(\vec{r})&\equiv&\frac{\vec{\tilde{\mathcal{P}}}_{\Omega_1,\Omega_2}^{(3)}}{\langle Q^{(2)}_{\Omega,\Omega}\rangle}\\
\nonumber
\tilde{Q}^{(3)}_{\Omega,\Omega,\delta\Omega}&\equiv&\frac{Q^{(3)}_{\Omega,\Omega,\delta\Omega}}{\langle Q^{(2)}_{\Omega,\Omega}\rangle},\\
\nonumber
\tilde{\vec{E}}^{(3)}_{\Omega,\Omega}(\vec{r})&\equiv&\frac{\vec{E}^{(3)}_{\Omega,\Omega}(\vec{r})}{\langle Q^{(2)}_{\Omega,\Omega}\rangle}.
\end{eqnarray}
With these definitions, $\tilde{Q}^{(3)}$ has dimensions of charge and $\tilde{E}^{(2)}$ and $\tilde{E}^{(3)}$ both have dimensions of charge per distance squared. The energy density is then more in line with $|\tilde{E}_{\Omega}^{(2)}(\vec{r})|^2/2$. Again, this would be, roughly, the energy density arising from the charges within $\Omega$. Correlators using any of the quantities scaled as above would then no longer grow with increasing $\Omega$ for systems with no long-range correlations. The charge correlators involving two operators labeled by $\delta\Omega$ would scale with $\delta\Omega$. 

Gauss's law relating  the scaled correlator $\tilde{E}^{(3)}$ to $\tilde{Q}^{(3)}$ is the same as that for the unscaled quantities.
\begin{eqnarray}
\oint_{\Omega} d\vec{A}\cdot\langle\vec{\tilde{E}}^{(3)}_{\Omega,\Omega}(\vec{r})\rangle&=&
\langle \tilde{Q}^{(3)}_{\Omega,\Omega,\Omega}\rangle,
\end{eqnarray}
or for arbitrary hyper-surfaces
\begin{eqnarray}\label{eq:gauss_scaled}
\oint_{\Omega} d\vec{A}\cdot\langle\vec{\tilde{E}}^{(3)}_{\Omega,\Omega}(\vec{r})\rangle&=&
\int d\Omega_{\mu}\cdot\langle \tilde{j}^{\mu(3)}_{\Omega,\Omega}(\vec{r})\rangle.
\end{eqnarray}

The Kubo relations for the scaled quantities become:
\begin{eqnarray}
\langle \vec{\tilde{j}}^{(3)}_{\Omega\Omega}(\vec{r})\rangle&=&\sigma \langle\tilde{\vec{E}}^{(3)}_{\Omega,\Omega}(\vec{r})\rangle,\\
\nonumber
\langle\tilde{\vec{\mathcal{P}}}_{\Omega,\Omega}^{(3)}(\vec{r})\rangle&=&\kappa \langle\tilde{\vec{E}}^{(3)}_{\Omega,\Omega}(\vec{r})\rangle.
\end{eqnarray}
The relations for baryon conductivity and polarizabilty become
\begin{eqnarray}\label{eq:kuboscaled}
\langle\vec{j}_B(\vec{r})\rangle&=&\frac{\langle B_{\delta\Omega}\tilde{Q}^{(3)}_{\Omega,\Omega,\delta\Omega}\rangle}
{\langle[\tilde{Q}^{(3)}_{\Omega,\Omega,\delta\Omega\rangle}]^2\rangle}
\sigma\langle\vec{\tilde{E}}^{(3)}_{\Omega,\Omega}(\vec{r})\rangle,\\
\nonumber
\langle\vec{P}_B(\vec{r})\rangle&=&\frac{\langle B_{\delta\Omega}\tilde{Q}^{(3)}_{\Omega,\Omega,\delta\Omega}\rangle}
{\langle[\tilde{Q}^{(3)}_{\Omega,\Omega,\delta\Omega\rangle}]^2\rangle}
\kappa\langle\vec{\tilde{E}}^{(3)}_{\Omega,\Omega}(\vec{r})\rangle.
\end{eqnarray}
The ratio of correlators in the Eq.s (\ref{eq:kuboscaled}) would then not grow with $\Omega$, approaching a constant. Thus, the fluctuation of the baryon current in the center of a long flux tube would be proportional to the fluctuation of $\tilde{E}^{(3)}$ which would not continue to grow with $\Omega$. 

\subsection{Applicability of Baryon Transport Relations}

The expressions in Eq. (\ref{eq:kuboscaled}) for the induced baryon current or polarizability provide insight into how color charge couples to baryon transport. For a system with non-zero conductivity, the application of the field $\vec{\tilde{E}}^{{(3)}}$ induces a current. Combined with Gauss's laws, Eq.s (\ref{eq:gauss}) and (\ref{eq:gauss_scaled}), one see how a sub-volume with quadratic charge $\tilde{Q}^{(2)}$ provides an energy per unit length to a tube of area $A$, in Eq. (\ref{eq:tension}). Further, if the sub-volume has a cubic charge, $\tilde{Q}$, it provides a field $E^{(3)}$, through Gauss's law. This field can then induce a baryon current, or if there is no conductivity, it induces a baryon polarization. 

The Gauss's law relations offer particularly nice insight into flux tubes due to the simple geometry. In that case one first chooses some point along the tube, and the subvolume $\Omega$ is then chosed as every part of a flux tube to one side of that point. By assuming that the field is constant across a cross-sectional area that divides a tube, one knows both the field, and thus the induced baryon transport. However, because this insight was based on Gauss's law, it does not necessarily translate well outside the paradigm of a flux tube. Even for a spherical volume one can imagine that the system has zero quadratic or cubic charge. But, that charge might have a dipole, or quadrupole moment. Thus, in general, it is challenging to calculate the field correlators outside the flux tube approximation. Even in the flux tube approximation, the Kubo relations are questionable because they are based on perturbation theory due to a slowly-changing applied field. However, in reality, such fields exist only in extremely dynamic systems, such as during the first half fm/$c$ of a hadronic collision. Even inside a hadron, the relations are difficult to motivate because the quarks and antiquarks are not isolated to one side of some volume. Finally, one should not forget that perturbation theory fails in QCD for any length or time scales $\gtrsim$ 1 fm.

Despite the limits to the applicability mentioned above, the relations provide useful insight, even if that insight only provides some theoretical backbone to heuristic arguments. In the next section, compound flux tubes, i.e. tubes where the sub-volumes are in states of higher-order color multiplets, are presented. This phenomenology might have more promise for further development, and perhaps provide predictive capability.

\section{Compound Flux Tubes}\label{sec:compound}
\begin{figure}
\centerline{\includegraphics[width=0.5\textwidth]{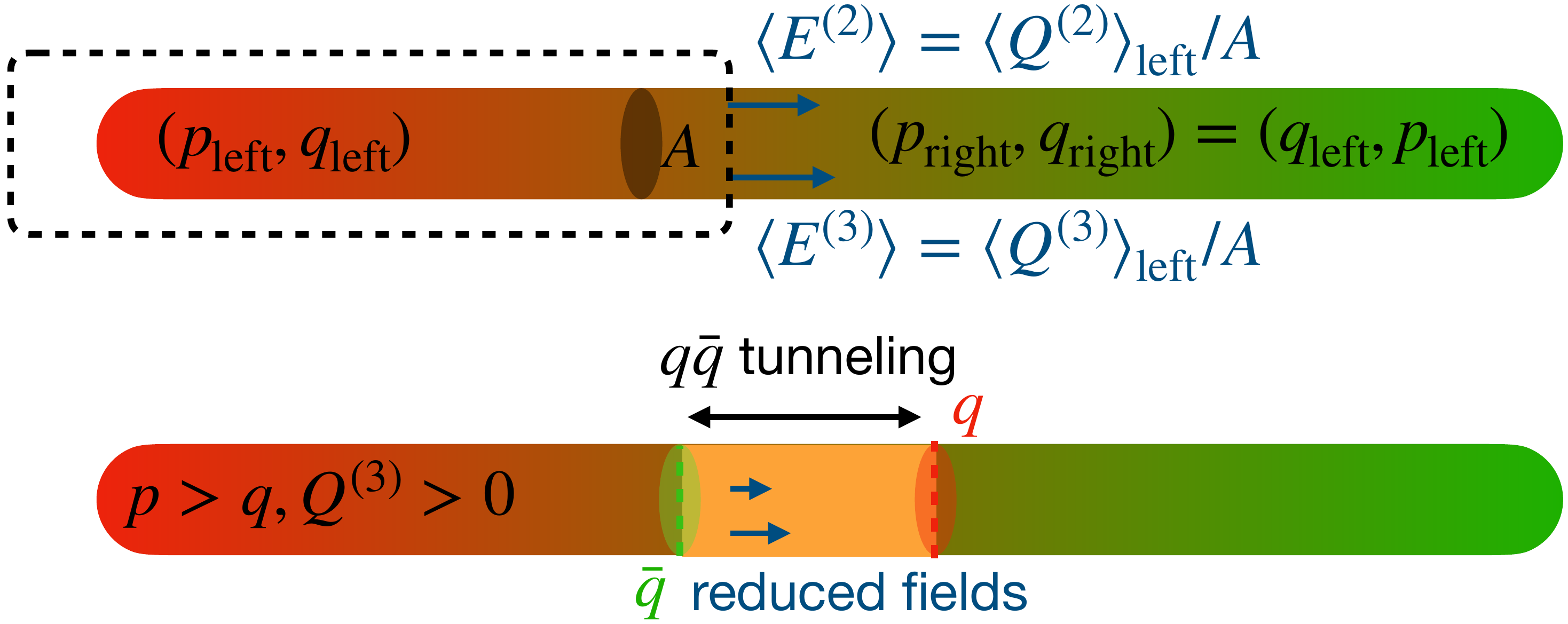}}
\caption{\label{fig:compoundtube}(color online)
The upper illustration show how a Gaussian surface isolates a flux tube into to parts. The color multiplets have interchanged values of $p$ and $q$. The energy density is defined by $\langle Q^{(2)}\rangle$, where the charge operators cover the left-hand side. 
}
\end{figure}

In Section \ref{sec:simple}, simple flux tubes were discussed. The term ``simple'' was a tube with a quark at one end and an antiquark at the other. At any dividing point, the color multiplet to one side was either $(p=1,q=0)$ or $(p=0,q=1)$. Here, we discuss tubes where at the dividing point the multiplet to each might be described by a higher multiplet. For example, moving a single gluon from one side to the other would result in either side of the tube being described by the octet $(p=1,q=1)$. By having multiple partons on a given side, any multiplet $(p,q)$ might be realized. The goal of this section is to investigate the possibilities of such combinations, and to understand how they might decay to the vacuum, which is a singlet $(p=0,q=0)$. The term ``compound'' tube refers to a tube with arbitrary $p$ and $q$. Similar considerations for more highly excited color flux tubes, or strings, have gone under the moniker of ``color ropes'' \cite{Sorge:1995dp,Bierlich:2014xba,Goswami:2019mta,Andersson:1991er}, and have been discussed as a means to increase nuclear stopping or enhance the production of heavy resonances.

Figure \ref{fig:compoundtube} illustrates a flux tube , where a Gaussian surface bisects the tube at some point as shown in the upper diagram. If the multiplet to one side is $(p,q)$ the multiplet describing the opposite side is $(q,p)$. Otherwise, the tube would not be in an overall color singlet. At the dividing point of the tube, the field-like color correlators are given by the color charges $Q^{(2)}$ and $Q^{(3)}$, which are simple functions of $p$ and $q$ defined in Eq. (\ref{eq:Q2Q3vspq}). The field energy is determined by $\langle Q^{(2)}\rangle_{\rm left}$ and the propensity to induce a baryon current is described by $\langle Q^{(3)}\rangle_{\rm left}$, where ``left'' refers to the fact that the volume to define the charge is the left side of the dividing point. If a $q\bar{q}$ pair tunnels through the vacuum, as shown in the lower diagram of Fig. \ref{fig:compoundtube}, and if one of those charges passes the dividing point, the color multiplet of matter to the left of that point changes. If it lowers $\langle Q^{(2)}\rangle_{\rm left}$, the energy density at the dividing point is decreased. Eventually, $q\bar{q}$ pairs must be produced in such a way that one reaches a color singlet with $\langle Q^{(2)}\rangle_{\rm left}=0$.

For the field-like correlators at some point to vanish, partons must pass that point, where the defining Gaussian surface intersects the flux tube. When a parton pass through the surface, the color state of the matter changes. If the initial multiplet is $(p_i,q_i)$ the available final states $(p_f,q_f)$ depend on whether the parton was a gluon, a quark or an antiquark.
\begin{eqnarray}\label{eq:pfqflist}
(p_i,q_i)+{\rm gluon}&\rightarrow& (p_f,q_f)=\left\{
\begin{array}{l}
(p_i+2,q_i-1)\\
(p_i+1,q_i-2)\\
(p_i-1,q_i+2)\\
(p_i-2,q_i+1)\\
(p_i+1,q_i+1)\\
(p_i-1,q_i-1)\\
(p_i,q_i)~~({\rm two~multiplets})
\end{array}
\right.\\
\nonumber
(p_i,q_i)+{\rm quark}&\rightarrow& (p_f,q_f)=\left\{
\begin{array}{l}
(p_i+1,q_i)\\
(p_i,q_i-1)\\
(p_i-1,q_i+1)
\end{array}
\right.\\
\nonumber
(p_i,q_i)+{\rm antiquark}&\rightarrow& (p_f,q_f)=\left\{\begin{array}{l}
(p_i,q_i+1)\\
(p_i-1,q_i)\\
(p_i+1,q_i-1)
\end{array}
\right.
\end{eqnarray}
\begin{figure}
\centerline{\includegraphics[width=0.75\textwidth]{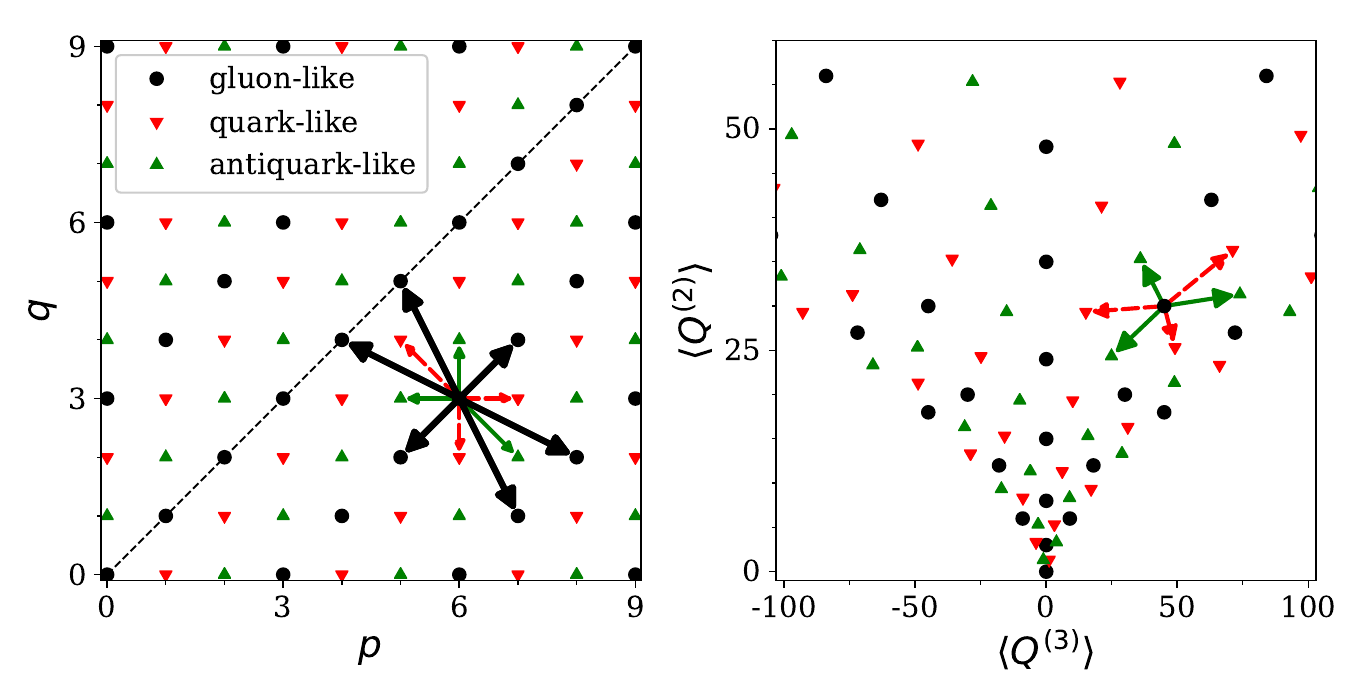}}
\caption{\label{fig:pq3}(color online)
Color multiplets are shown as a function of $(p,q)$ in the left-side panel. Starting from the singlet, ($p=0,q=0$), one can combine gluons to reach any of the multiplets denoted by black circles. These ``gluon-like'' states,  can also be reached by combining quarks and antiquarks if the difference of the number of quarks and antiquarks is a multiple of three. To reach a ``quark-like state'', one where $(p-q)/3$ has a remainder of +1 or -2, one can add a quark to a gluon-like state. One can reach an ``antiquark-like state'', one where the remainder of $(p-q)/3$ is -1 or 2, by adding an antiquark to a gluon-like state. The heavy black lines show the states available after adding either a gluon (heavy black line), an antiquark (light green line) or a quark (light dotted line). If gluon creation is possible, the most efficient way the system can reach a color singlet is to create gluon pairs, or perhaps one quark or antiquark depending on whether the original state is gluon-like, quark-like or antiquark-like. If gluonic steps are not permitted the most efficient decay mechanism is through $p$ quarks and $q$ antiquarks. The right-side panel represents the same states, but the states are plotted as $\langle Q^{(2)}\rangle$ vs $\langle Q^{(3)}\rangle$. In this basis the energy per unit length is a function of $\langle Q^{(2)}\rangle$ and the field driving the baryon current is $\langle Q^{(3)}\rangle$.}
\end{figure}

If the initial tube is created by exchanging only gluons across the dividing point, the available multiplets must have $p_f-q_f$ in multiples of three. Even though such states were created by gluons passing through, both $p_f-q_f$ and $\langle Q^{(3)}\rangle_{\rm left}$ can be non-zero. Thus, they can induce baryon transport across the dividing point. However, if the decay involves gluon pairs, one can also return to the ground state without any quarks passing the dividing line. If quarks and or antiquarks were initially exchanged across the dividing line, one can also return to the ground state through the creation of gluon pairs if the difference in the quark number is a multiple of three. One can define a quark excess number as
\begin{eqnarray}
B_x&=\left\{\begin{array}{rl}
1,&N_{\rm quarks}-N_{\rm antiquarks}=\cdots,-5,-2,1,4,7\cdots\\
-1,&N_{\rm quarks}-N_{\rm antiquarks}=\cdots,-7,-4,-1,2,5\cdots\\
0, &N_{\rm quarks}-N_{\rm antiquarks}=\cdots,-6,-3,0,3,6\cdots
\end{array}\right.
\end{eqnarray}
If $B_x=1$ the multiplet is ``quark-like'', and at least one antiquark must pass into the left-hand side of the tube, or a quark must pass from left to right. Similarly, if $B_x=-1$ the multiplet is ``antiquark-like'', and at least one quark must pass the dividing point from right to left, or an antiquark must pass from left to right. Additional groups of three quarks and three antiquarks are possible. This was expected, knowing that the net quark number on each side must be a multiple of three in the final state. If gluon pair creation is allowed, it represents the most efficient path to decaying the state in terms of the number of parton pairs created. In that case the net baryon number that passes the dividing point is $-1/3,1/3$ or zero.

If during the pair creation process the ability to create gluon pairs becomes negligible, returning to a color singlet requires $q-\bar{q}$ pair creation. The states available to each step are listed in Eq. (\ref{eq:pfqflist}) and are illustrated in Fig. \ref{fig:pq3}. If one couples a large number of partons, one can create multiplets with large value of $p$ and $q$. Multiplets with $p\approx q$ should be more likely because there are more ways in the creation process to couple a given number of partons to such a multiplet. Further, the degeneracy of a give multiplet, $D(p,q)=(p+1)(q+1)(p+q+2)/2$, increases both for higher $p+q$ and for lower $|p-q|$. Nonetheless, there is the possibility of creating multiplets with $p$ and $q$ significantly differing.

Because adding a quark (or antiquark) into a region during the decay process can lead to three different multiplets, there are numerous possible paths from the original $(p_i,q_i)$ that lead to a color singlet. The most efficient path, that with the fewest partons passing the dividing point, is one where $p_i$ antiquarks and $q_i$ quarks cross into the left side, or if $p_i$ quarks or $q_i$ antiquarks move to the right side. Thus, if through multiple-gluon exchange the initial state has $p_i-q_i=6$, and if the breakup stage strongly discourages the creation of gluon pairs compared to $q\bar{q}$ pairs, two baryonic charges, or six quarks, would move across the dividing point. 

Should the latter possibility, that gluon pair creation is discouraged during the decay process, be valid, one could imagine that baryon number might have to cross a large swath of rapidity for the system to recombine into color singlets. This does not, however, imply that a single baryon moved across the fluxtube. Instead, a large number of polarized $q\bar{q}$ pairs may have appeared. Just as polarizing a dielectric results in a surplus of electric charge on one side and a dearth of charge on the opposite side, these pairs appear without any actual movement of charge aside from the charges in individual pairs moving a small distance apart from one another to create the dipole density.

In this case, where the tubes decay only via the tunneling of $q\bar{q}$ pairs, the number of quarks moving by a certain point is defined by the color multiplet, or more specifically by $p-q$, on either side of the point. The end of each tube is ultimately defined by a quark, antiquark or gluon, so the value of $p-q$ at the end of the tube is small. At each point in the tube where a gluon was placed in the original state, the running value of $p-q$ changes, and the number of quarks or antiquarks passing these points also changes. Thus, at these points baryon number accumulates. For that reason if one thought of a tube as being orignally occupied by some number of partons spread along the tube, the decay process would likely lead to baryons or antibaryons congregating at those points. If one were to plot $p-q$ of the initial state vs. the position or spatial rapidity along the tube, $\eta$, and if the initial state was driven purely by gluons occupying various positions along the tube, the values of $p-q$ would either rise or fall by three, or stay the same, at each gluon's position. After the $q\bar{q}$ decays, antibaryons would appear at those positions where $p-q$ increased by three, and baryons would appear at those positions where $p-q$ decreased by three. 

\begin{figure}
\centerline{\includegraphics[width=0.8\textwidth]{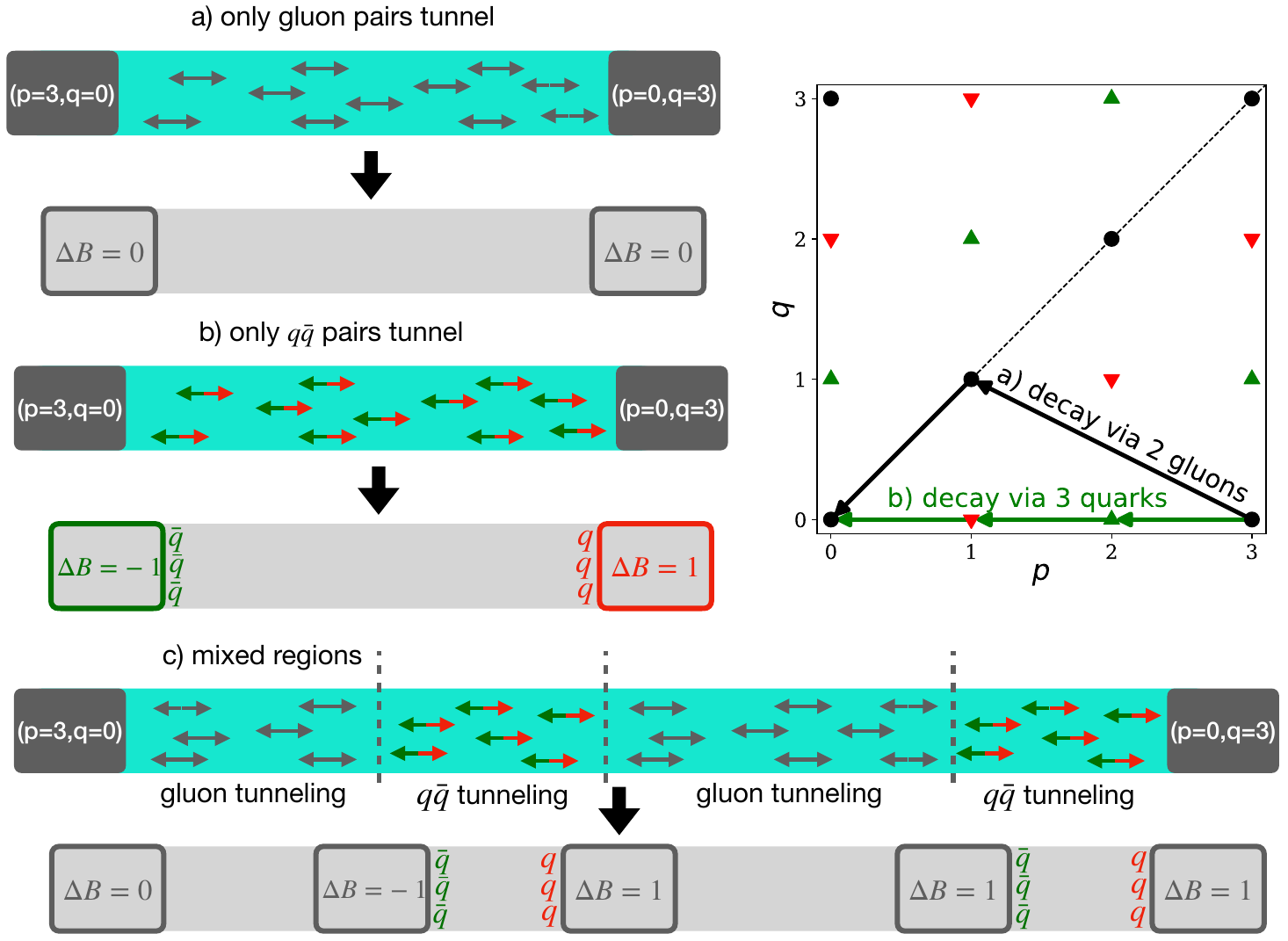}}
\caption{\label{fig:2gluon}(color online)
An illustrative example of how a flux tube, whose initial state is characterized by a multiplet $(p=3,q=0)$ to the left, might decay. In the path (a), for any point along the tube, two gluons pass by. This restores the states to the left and right of that point to color singlets. The first gluon alters the left-side (as defined from the point) from $(p=3,q=0)$ to $(p=1,q=1)$. The right side also becomes a color octet. The second gluon brings both sides to a color singlet. No baryon number coalesces in this scenario. For path (b), only quarks pass any given point. This requires some combination of three quarks or antiquarks to pass in order to create color singlets. Either three antiquarks pass the point towards the left, three quarks pass to the right, two quarks to the right and one antiquark to the left, or one quark to right and two antiquarks to the left. For any such change, the final state has an antibaryon on the left and a baryon to the right. Finally, in (c) an example is shown where in one region, three quarks pass by, while in another, two gluons pass. In this case, the final configuration has a baryon and antibaryon located at the opposite ends of the region where quarks and antiquarks were responsible for the $j^{(3)}$ current.
}
\end{figure}

Figure \ref{fig:2gluon} provides an illustration of how a tube might decay if the tube is driven by two charge multiplets, $(p=3,q=0)$ on the left edge and $(p=0,q=3)$ on the right edge.  These states might have been created by two gluons being exchanged between some target and some projectile. Assuming that the intermediate region is void of charge, the field-like correlators evaluated at some point in the intermediate region are defined by $p$ and $q$. For any position along the tube, the field energy will not disappear until come color charges have moved across that position. Then, the matter to the left, or right, of the specified position will then both be in color singlets. The number of quanta must be two or more. Two gluons crossing the position can accomplish the feat, as pictured in the upper illustration of Fig. \ref{fig:2gluon}. The first gluon can change each side to the multiplet $(p=1,q=1)$. The second gluon can then transform the sides into color singlets. If gluon pairs are readily produced from the field, or if gluons can be easily exchanged across the position, this should be the dominant mechanism for decay. No baryon current would be generated and no baryon anti-baryon pairs would be produced. Should gluonic modes not be easily excited, the decay of the tube would rely on the creation of $q-\bar{q}$ pairs. In that scenario, the most likely means of decay would require three quarks, or antiquarks, to pass any pre-specified position. As long as the net baryon charge moving to the left side is -1, one can divide the tube into two tubes, each in a color singlet. As shown in Fig. \ref{fig:2gluon} one side of the tube would have a baryon, while an anti-baryon would populate the opposite side. Of course, it is possible that both gluonic and $q\bar{q}$ decay modes might contribute. In that case, some positions on the tube might see two gluons cross in order to return both sides to a color singlet, while a different position might observe quarks passing through. In that case, the regions with quark movement would distill a baryon anti-baryon pair spanning the region. Other decay routes are also possible For example, a gluon might bring the system to the state $(p=1,q=1)$, followed by a quark and antiquark moving in the same direction, two quarks passing in opposite directions, or two antiquarks passing in opposite directions. No baryon charge would be distilled in this scenario.

From viewing Fig. \ref{fig:2gluon} one can see how a flux tube, even one created purely from a gluonic excitation, might decay is such a way to not only create baryon/anti-baryon pairs, but to significantly separate the baryon charge in rapidity. The lower illustration in Fig. \ref{fig:2gluon} shows an example where the decay of a $(p=3,q=0)$ state might be proceeding through gluonic transport in some segments of the tube, while proceeding via quark-antiquark transport in other segments. Assuming there is some characteristic length scale, $\lambda$, over which one cannot change from one decay mode to the other, and assuming that the probability of being in the $q\bar{q}$ mode at any given point is $P_{q\bar{q}}$, the probability of having $N_{q\bar{q}}$  adjacent steps of size $\lambda$ all in the $q\bar{q}$ mode would be $P_{q\bar{q}}^{N_{q\bar{q}}}$, and the distributions of $q\bar{q}$ region sizes would fall off exponentially as a function of the distance $L$, i.e. $\propto \exp\{-(\ln(P_{q\bar{q}})L/\lambda\}$. The separation of balancing baryon/anti-baryon pairs would then be characterized by an exponential falloff. For the $(p=3,q=0)$ example illustrated in Fig. \ref{fig:2gluon} one could imagine that the color multiplet on the left side involved a baryon, which emitted two gluons to the right side. In that case, the three quarks could join with the three antiquarks distilling from the color field to produce mesons. The end result would be that the baryon number at the left would move to the boundary of the $q\bar{q}$ like decaying region. Thus, the baryon movement would also fall off exponentially, similar to the behavior expected from the simple flux tubes in Sec. \ref{sec:simple}, which involved merging tubes. However, even though the two process of baryon transport, one from a compound tube and the other from the merging of simple tubes, should both transport baryons from beam or target rapidities with an exponential falloff, there is no reason to think the two exponential scales would be the same.

Again, the mechanism for baryon-antibaryon separation would be  weaker for electric charge separation, vanishing in the limit that $u\bar{u},~d\bar{d}$ and $s\bar{s}$ tunneling became equally possible. Separation of balancing charges has indeed been measured, and has even been binned by whether the balancing charges are pions, kaons or protons. Indeed, the separation from protons has been observed to be significantly longer than that from pions in heavy ion collisions. However, much of that difference may simply arise from the fact that much of the electric charge is created late in the collision, through hadronization and decays. It would be interesting to experimentally analyze this class of correlations, known as charge balance functions, for $pp$ or $p\bar{p}$ collisions. Most current analyses of charge-balance functions have centered on relative rapidities, $\Delta y \lesssim 1$, e.g. \cite{Pratt:2021xvg}. For the purposes here, it would be helpful to consider measurements of baryon-antibaryon correlations at somewhat larger relative rapidities.

\section{Gluon Radiation in the Intermediate Region}\label{sec:gluonradiation}

In the first section, where only simple flux tubes were considered, the excited tube was created via the exchange of a gluon between the target and projectile valence quarks. Even if multiple gluons are emitted from a baryon, the multiplets of any three-quark states would be confined to $(p=0,q=0)$, $(p=1,q=1)$ or $(p=3,q=0)$. To maintain an overall color singlet for two nucleons, the correpsonding projectile states would have to be $(p=0,q=0)$, or $(p=1,q=1)$. The multiplets $(p=3,q=0)$ and $(p=0,q=3)$ can couple to a color singlet, but three quarks cannot carry the $(p=0,q=3)$ possibility. Thus, if gluons are only ``exchanged'', the only possible non-zero color flux tube for nucleon-nucleon collisions is the $(p=1,q=1)$ octet. If the projectile were an antinucleon, one could also have $(p=3,q=0)$ excitation of the color flux tube. Assuming that a single flux tube connects the target and projectile, the energy density of such a tube is uniform along the tube. However, if the gluons are not only exchanged, but radiated, one an augment this picture by placing gluons along the flux tube. This enables higher color excitations. One might traverse several gluons between the ends of the tube, where the three quarks reside, to mid-rapidity. The goal of this section is discuss the implications of adding intermediate gluonic color charges to the flux tube picture. In addition to the impact of these charges toward baryon transport, these charges are shown to modify the multiplicity density distributions as a function of rapidity. 

\begin{figure}
\centerline{\includegraphics[width=0.45\textwidth]{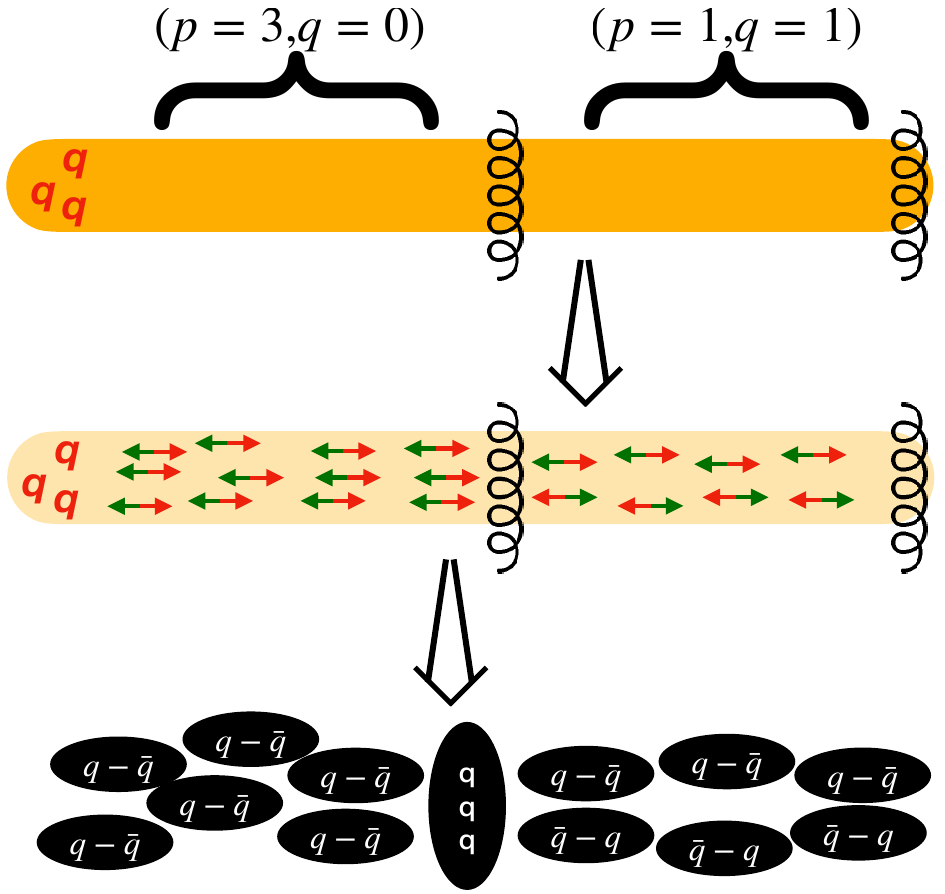}}
\caption{\label{fig:gluon_radiation}
A compound flux tube driven at one end by three quarks might create a tube by radiating two gluons. As one travels along the tube from the left until the nearest emitted gluon is encountered, the region might be characterized by $(p=3,q=0)$. If this state hadronizes by pair creation, the baryon number on the left will be absorbed by the polarizing pairs, and in effect, transport the baryon number to the position of the first gluon, as illustrated in the lower part of the illustration. The remaining segment can hadronize by creating quark-antiquark pairs where there is no baryon polarization, e.g. one could have two simple flux tubes with opposite directions of baryon polarization. This illustrates how radiated gluons can lead to flux segments characterized by $p\ne q$, which then result in baryon transport.
}
\end{figure}

Figure \ref{fig:gluon_radiation} illustrates a compound tube with three quarks on one end, where two gluons are radiated along the tube. The quarks might find themselves in any of three multiplets: $(p=0,q=0),(p=1,q=1)$ or $(p=3,q=0)$. As an example, Fig. \ref{fig:gluon_radiation} displays the case where the quarks are in a $(p=3,q=0)$ state. In that event, if one travels down the tube, the parameters of the multiplet can change when one encounters a gluon, which in the illustrated case couples to $(p=1,q=1)$. For the matter to the left of the gluon's position to become color singlets via quark-antiquark pair creation, the polarization must result in the baryon number moving to the position of the gluon. At that point the three left-over quarks can couple with the gluon to form a baryon, plus an additional gluon that would be part of the tube to the right. Thus, this example illustrates how baryon number can be transported along a compound flux tube to the position of a radiated gluon. Again, this is predicated on the color fields neutralizing via quark-antiquark production, rather than gluon-pair creation.

In this section, we consider the possibilities created by emitting a number of gluons, $N_g$, along the flux tube. It is assumed that the flux begins and ends with an integral number of baryons, and that all the gluons are located along the tube between the quarks. An additional ansatz is applied, that the color states are originally determined randomly, aside from the constraint that the system is in an overall color singlet.

\subsection{Randomly Colored States of Quarks}

Before considering an entire flux tube, we first consider the color state of $N_q=3B$ quarks at one end of a tube. To get an idea how the population of color states might behave in the limit that many gluons are radiated, an ensemble of randomly colored quarks is considered. Of particular interest is determining whether having $B>1$ might lead to increasingly high baryon transport. For example if three quarks are at one end of the tube, the color states available are $(p,q)=(0,0),(1,1)$ or $(3,0)$. Whereas the first two multiplets have $p=q$, the last one has $p-q=3$. If one randomly averages over the three multiplets weighted by their degeneracy, the average of $p-q$ is $\langle p-q\rangle=30/19$. One might ask what the $\langle p-q\rangle$ would be if had four baryons, 12 quarks, and if their color state was assigned randomly. A motivation for considering such a compound flux tube where one end has $B>1$ comes from $p-A$ collisions or from heavy-ion collisions. In that case one might think of the system as having many tubes forced to merge into a single tube.  For example, if $\langle p-q\rangle$ increases by a factor of four, one would expect four times the baryon transport because the baryon transport should roughly scale as $p-q$.

\begin{figure}
\centerline{\includegraphics[width=0.5\textwidth]{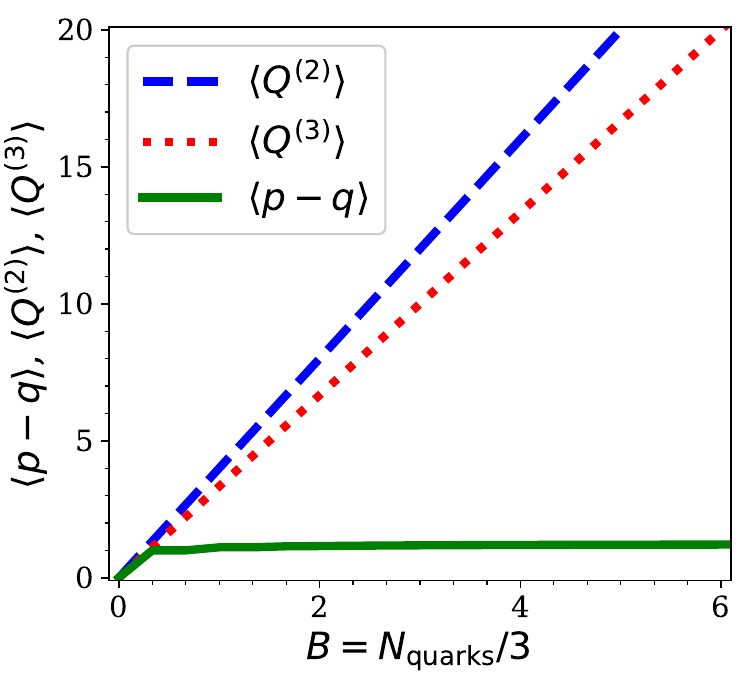}}
\caption{\label{fig:pminusq}
$N_q=3B$ quarks are randomly coupled in color space, the average of the quadratic Casimir charge, $\langle Q^{(2)}\rangle$, and the average of the cubic Casimir charge, $\langle Q^{(3)}\rangle$, grow linearly with the number of quarks being coupled. However, $\langle p-q\rangle$ rapidly plateaus. Because it is $\langle p-q\rangle$ that is most correlated with the transport of baryons, having many quarks at one end of the tube, with their colors assigned randomly, should not induce large baryon currents. In contrast, because $\langle Q^{(2)}\rangle$ rises linearly with $B$, overlapping flux tubes from individual baryons will result in similar energy density as having all the baryons contribute individually.
}
\end{figure}

With that motivation, color states are generated by iteratively coupling $N_q$ quarks to $N_q+1$ quarks. As described in Eq. \ref{eq:pfqflist}, three new multiplets for $N_q+1$ can be created based on any $N_q$ multiplet. The new multiplet is chosen randomly, but weighted according to the degeneracy of the possible new multiplets. The number of color states, $3^{N_q}$, should equal the sum of each produced multiplet multiplied by its degeneracy. Some values of $(p,q)$ might appear multiple times. Figure \ref{fig:pminusq} shows the resulting average values, $\langle Q^{(2)}\rangle, \langle Q^{(3)}\rangle$ and $\langle p-q\rangle$, as a function of $B$, where the averaging is over all the possible degneracy-weighted multiplets. The two quantities $\langle Q^{(2)}\rangle$ and  $\langle Q^{(3)}\rangle$ behave perfectly linearly,
\begin{eqnarray}
\langle Q^{(2)}\rangle&=&4B,\\
\langle Q^{(3)}\rangle&=&\frac{10}{3}B.
\end{eqnarray}
Because the energy per unit length of the tube grows roughly linearly with $Q^{(2)}$, one can see that adding more baryons increases the energy density of the tube as if the baryons were contributing independently if one averages over all color multiplets. The average value of $Q^{(3)}$ also rises linearly, but it is $p-q$, rather than $Q^{(3)}$, that determines the number of baryons likely to pass a given point. Given that $\langle p-q\rangle$ quickly saturates with $B$, one would not expect much baryon transport to occur once tubes have merged. This suggests that in a heavy-ion environment most of the transport of baryons from the target or projectile should take place before, or during, the merging of the flux tubes from individual nucleons.  

\subsection{Random Color Ansatz (RCA) for Color Flux Tubes with Intermediate Gluons}

We consider a simple picture of a flux tube, where $N_{\rm proj}=3B_{\rm proj}$ quarks from the projectile are concentrated on one end of the tube, while $N_{\rm target}=3B_{\rm target}$ quarks are placed at the other end. Additionally, $N_g$ gluons are placed along the tube. The rapidities of the projectile and target are scaled to the variable $\tilde{y}$, with $\tilde{y}=\pm 1$ referring to the target and projectile respectively. For illustration purposes, the gluons are placed at equal intervals between $\tilde{y}=\pm 1$. One could easily adjust the positions by simply remapping to represent a different distribution of gluons in rapidity. Trajectories for each segment are created from the left to the right by simply coupling in a new quark or gluon as it is passed, with all couplings given equal probability, aside from the color degeneracy factor, $D(p,q)=(p+1)(q+1)(p+q+2)/2$. Once the last quark is coupled, any trajectory that did not result in a color singlet is discarded. 

\begin{figure}
\centerline{\includegraphics[width=0.5\textwidth]{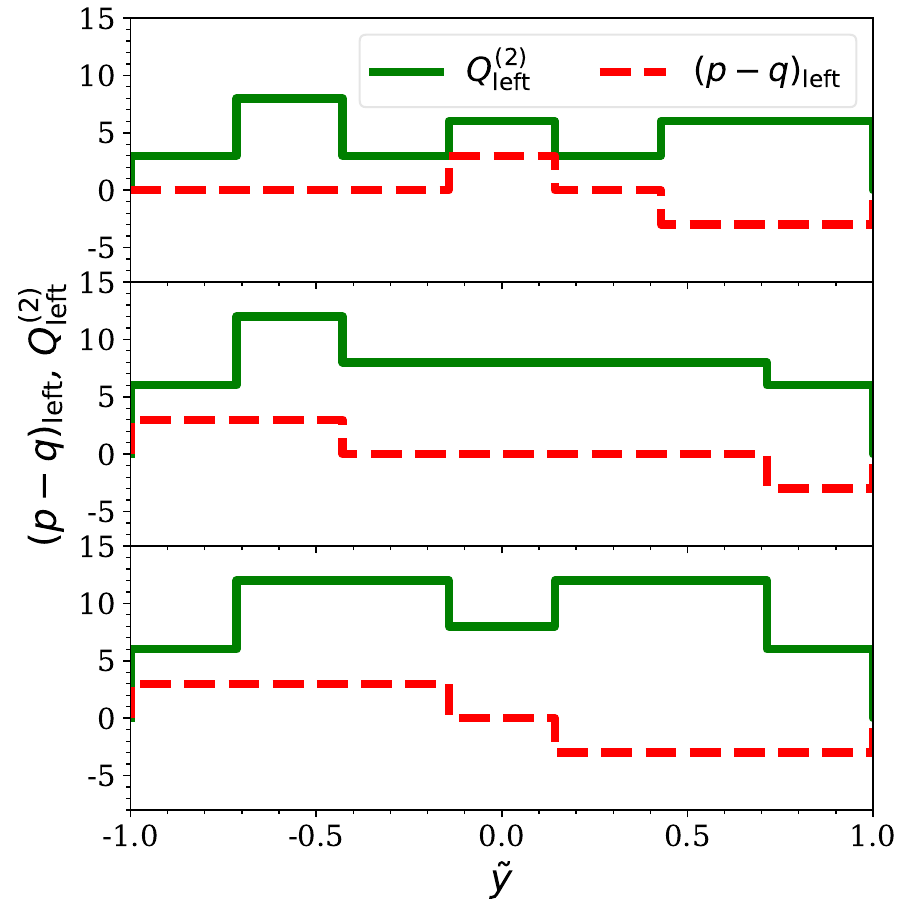}}
\caption{\label{fig:sampletrajectories}
For the random color ansatz, three trajectories in color space are displayed as a function of the scaled rapidity $\tilde{y}$. The trajectories all begin and end as color singlets. The intermediate segments have fluctuating values of $Q^{(2)}_{\rm left}$, which corresponds to fluctuations of the energy density, and fluctuating values of $p-q$, which drive fluctuating baryon currents.}
\end{figure}

Figure \ref{fig:sampletrajectories} displays three such trajectories. Each trajectory was created with $B_{\rm target}=B_{\rm proj}=1, N_g=6$. Each segment of the tube is characterized by the labels, $(p,q)$, or equivalently $Q^{(2)}_{\rm left},Q^{(3)}_{\rm left}$. The energy density, or the string tension, of the tube is proportional $Q^{(2)}_{\rm left}$, and the baryon transport is driven by $p-q$. If one had defined the labels based on the right-hand side of the segment, the labels $p$ and $q$ would be reversed. This would leave $Q^{(2)}_{\rm left}$ unchanged, and would flip $Q^{(3)}_{\rm left}$. 

\begin{figure}
\centerline{\includegraphics[width=0.95\textwidth]{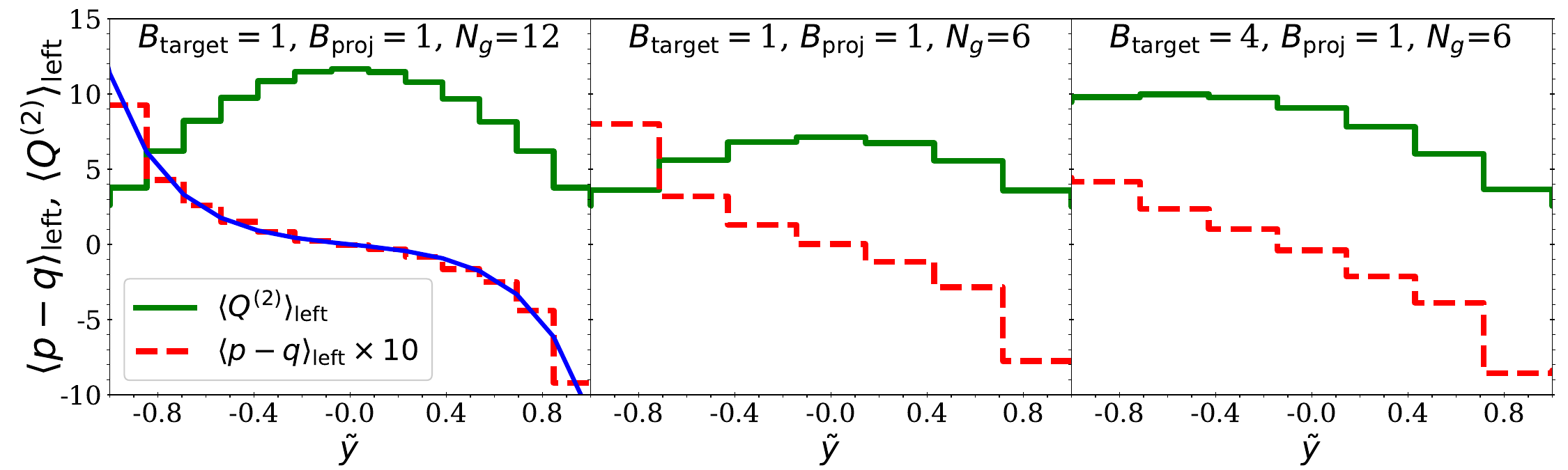}}
\caption{\label{fig:averagetrajectory}
Values of $p$ and $q$ were averaged over many trajectories for three cases. For the first two cases, $B_{\rm target}=B_{\rm proj}=1$ and $N_g=$ 6 or 12. The energy density, whose proxy is $Q^{(2)}_{\rm left}$, rises at mid-rapidity, with a stronger $\tilde{y}$ dependence for higher $N_g$. This can be thought of as a random walk in color space over $N_g$ steps, where the walk is constrained to begin and end at $(p=0,q=0)$. The greatest distance from the origin will occur at $\tilde{y}=0$ and will increase with the number of steps, i.e. $N_g$. Thus, the maximum of $Q^{(2)}_{\rm left}$ for $N_g=12$ is roughly twice that for $N_g=6$, which can be seen by comparing the left-side and center panels. The right-side panel illustrates the asymmetric case, with four baryons on one end and only one baryon on the opposite end. In that case the energy density is asymmetric. 
}
\end{figure}

Figure \ref{fig:averagetrajectory} summarizes the average trajectory of the random walk in color space for three cases. The first case, displayed in the central panel, is the same as used for the sample trajectories displayed in Fig. \ref{fig:sampletrajectories}, with $B_{\rm target}=B_{\rm proj}=1$ and $N_g=6$. The second case, first panel, differs by increasing the number of intermediate gluons, $N_g$, to 12. The blue line represents a fit to the form $A\sinh(\tilde{y})$. This fit is consistent with the baryon transport decaying exponentially from the ends of the tube. The third case increases $B_{\rm target}$ to 4 nucleons, while keeping $N_g=6$. The average of $\langle Q^{(2)}\rangle_{\rm left}$ can be thought of as a measure of the energy per unit length of the flux tube. For the two cases where $B_{\rm target}=B_{\rm proj}$, this is perfectly symmetric in $\tilde{y}$. Also shown in this panel is a fit of $Q^{(3)}_{\rm left}$,
\begin{eqnarray}
Q^{(3)}_{\rm left}\approx A\sinh(\tilde{y}/a),
\end{eqnarray}
where $a$ and $A$ are fit parameters. Because the baryon current would be generated proportional to $Q^{(3)}$, and because the baryon density should be proportional to the divergence of the current, the resulting baryon density would then be $\sim \cosh(\tilde{y}/a)$. The quality of the fit is  consistent with the baryon transport from the ends decaying exponentially, but with the caveat that this is the scaled rapidity $\tilde{y}$, not the real rapidity.
The right-hand panel of Fig. \ref{fig:averagetrajectory} displays the average trajectory for the case where one side of the string has four baryons and the opposite side has one, which might be reasonable for a central $p-A$ collisions. In this case the energy density at the four-baryon side starts higher because the 12 quarks can accommodate a higher color state. The distribution is then weighted toward that side. Emission is stronger at $\tilde{y}=0$ for the two symmetric cases. For the asymmetric case the multiplicity distribution is weighted toward the side of the tube where the nucleus is located. Both such features are seen experimentally \cite{BRAHMS:2003wwg,PHOBOS:2010eyu,ALICE:2016fbt,ALICE:2013jfw,ALICE:2022imr}. If one were to reduce the number of intermediate gluons the asymmetry would be more pronounced.

\subsection{Baryon-antibaryon Correlations}

For any produced baryon, one must produce an antibaryon to balance the baryonic charge. The quark-antiquark pairs that are responsible for baryon transport should also manifest themselves in what is known at the baryon balance function \cite{Pratt:2021xvg}. This observable identifies the location of balancing charge through a like-sign subtraction. For the random color ansatz described above, quark-antiquark pairs are created in any region with $Q^{(3)}\ne 0$ or $p-q\ne 0$. A current ensues with baryonic charge collecting at each location of the gluonic charges at which $p-q$. changed. If the color tubes decay exclusively via quark-antiquark production, the baryon number that appears at each gluonic position, $i$, is 
\begin{eqnarray}
 b_i&=&3(p-q)_{i-1,i}-3(p-q)_{i,i+1}.
 \end{eqnarray}
 Here, $(p-q)_{i,i+1}$ is the initial value of $(p-q)_{\rm left}$ in the interval between the $i^{\rm th}$ and $(i+1)^{\rm th}$ intermediate gluon. Because only one gluon is at a given position, $b_i$ must equal -1,0 or +1. The balance function of interest is
\begin{eqnarray}
B(\Delta N_g)&=&-\frac{\sum_{i\ne j}b_ib_j}{\sum_i|b_i|}\delta_{\Delta N_g,|i-j|}.
\end{eqnarray}
This represents the probability that for any baryonic charge $b_1$, one finds an additional oppositely charged baryon (relative to the same sign) at a position of $\Delta N_g$ relative to $b_1$. If the net baryon number is zero, this then integrates to unity. If the net baryon number is non-zero, one can modify the definition of $B(\Delta N_g)$ by replacing $b_i$ with $\delta b_i=b_i-\langle b_i\rangle$ to maintain the constraint. 

\begin{figure}
\centerline{\includegraphics[width=0.48\textwidth]{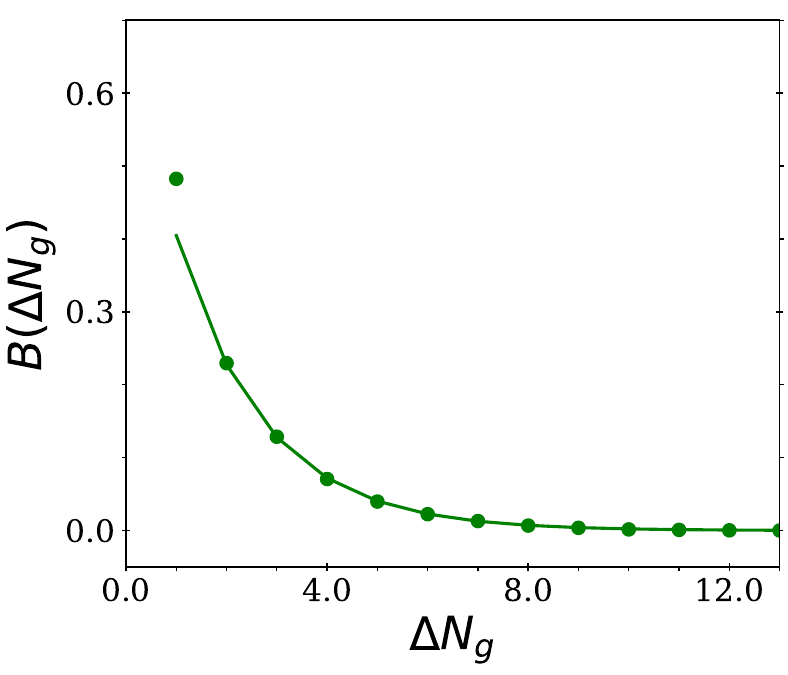}}
\caption{\label{fig:bcorr}
The baryonic balance function as applied to the random color ansatz, with 16 intermediate gluons and no quarks. For any observed baryonic charge, the balancing charge is more readily found in a nearby site, i.e. small $\Delta N_g$. The balance function, which sums to unity, falls off exponentially. The solid line shows a fit to the tail which behaves $\sim e^{-\alpha\Delta N_g}$, with $\alpha=0.576$. This shows that balancing baryonic charge is localized, and that the rapidity separation is roughly the rapidity separation between emitted gluons divided by $\alpha$.
}
\end{figure}
In order to isolate the physics of baryon transport from the effects of the ends of the tube, and to focus on the correlations of baryonic charge due to pair creation, the RCA is applied here with zero quarks at the end of the tube, and a large number of intermediate gluons, $N_g=16$. Using $\approx 2\times 10^5$ instances, the baryonic balance function was calculated using all the baryons and is displayed in Fig. \ref{fig:bcorr}. Aside from a small deviation at small $\Delta N_g$, the correlation falls off exponentially with $\Delta N_g$. The exponential fall off behaves approximately $\sim e^{-\alpha \Delta N_g}$, with $\alpha=0.576$, for the case with 16 gluons.

The extracted value of $\alpha$ falls if the number of gluons is increased. To understand this behavior one can consider a system with large $p+q$, but small $p-q$. The degeneracy can be expressed as
\begin{eqnarray}
D(p,q)&=&\frac{1}{8}\left[(p+q)^2-(p-q)^2+4(p+q+1)\right](p+q+2).
\end{eqnarray}
If one is in a state $p,q$ and adds a gluon, the preferred path will wish to reduce $|p-q|$ due to the degeneracy weight. The average change in $(p,q)$ can be found by averaging over the 8 different possible multiplets described in Eq. (\ref{eq:pfqflist}). 
\begin{eqnarray}
\langle \Delta(p-q)\rangle &=&\frac{1}{\sum_jD(p_j,q_j)}\sum_jD(p_j,q_j)(p_j-p-q_j+q),
\end{eqnarray}
where $j$ refers to each of the 8 multiplets. Taking the differential limit, which should be valid for large $p$ and $q$, and small $p-q$, 
\begin{eqnarray}\label{eq:pminusqfalloff}
\langle \Delta(p-q)\rangle &=&\frac{1}{\sum_jD(p_j,q_j)}\sum_jD(p_j,q_j)(p_j-p-q_j+q)\\
\nonumber
&\approx&\frac{1}{8D(p,q)}\frac{dD}{d(p-q)}\sum_j (p_j-p-q_j+q)^2\\
\nonumber
&=&\frac{9}{2D(p,q)}\frac{dD}{d(p-q)}\\
\nonumber
&\approx&\frac{-9}{(p+q)^2+4(p+q+1)}(p-q).
\end{eqnarray}
Since the change of $p-q$ is proportional to $p-q$, the falloff is exponential and $\alpha$ is the prefactor in the last line of Eq. (\ref{eq:pminusqfalloff}). This form explains both the exponential behavior and the dependence of $\alpha$ on the number of gluons, as the characteristic value of $(p+q)^2$ increase linearly with $N_g$.

If the gluons were distributed uniformly in rapidity, the separation of balance function for baryonic charge would also fall off exponentially in relative rapidity. The characteristic scale in relative rapidity would then be $\Delta y_g/\alpha$, where $\Delta y_g$ would be the typical separation between neighboring gluons in the initial flux tube. Because a given quark-antiquark pair would likely separate by $\Delta y_g$, the baryonic separation would likely be larger, by a factor of $1/\alpha$, than that for other balancing charges, such as strangeness or electric charge.

Balance functions for protons and antiprotons have been measured by the ALICE Collaboration for heavy ion collisions at several centralities \cite{ALICE:2021hjb}. Indeed, the widths are larger than what can be explained without having significant separation at the earliest times \cite{Pratt:2021xvg}. However, the data are confined to a tight rapidity window, which forbids accurate extraction of the long-range part of the correlation. Given that some pair creation might be of a different source, e.g. chemical thermalization of the energy, a measurement of baryon-antibaryon correlations at larger relative rapidity would be necessary to clarify the role played by the separation of quark-antiquark pairs pulled apart by longitudinal color fields. 

\section{Summary}
The previous sections focused on separate goals, but all were related to understanding how color flux is related to baryon transport and stopping. Sec. \ref{sec:simple} restated the ideas of \cite{Vance:1997th} and demonstrated how baryon number would be transported from either the projectile or target toward mid-rapidity due to the polarization of quark-antiquark pairs in simple flux tubes. It was shown that baryon number was transferred to the point where two simple tubes merged. Assuming the penalty for tubes not merging into one was proportional to the length of the region with two tubes, one would expect an exponential fall off of the transport with relative rapidity. Further, because the polarization should be approximately equal for $u\bar{u},~d\bar{d}$ and $s\bar{s}$ pairs, there should be little transport of electric charge. The lack of a quantitative scale for the exponential falloff prevents these arguments from being completely convincing. Nonetheless, this explanation, which is based on a rather standard picture of hadronization through pair creation, does remove any sense of puzzle from the experimental observation.

The second section is more formal in nature, and provides an understanding of how color electric fields can lead to baryon transport. It was found that one can define quantities that play the roles of chargess and fields that are related to the cubic Casimir of SU(3). Relations that parallel typical expressions for linear response are derived. The perturbative nature of these relations is contradicted by the clear understanding that the phenomenology of color flux tubes is explicitly non-perturbative. But, despite this inconsistency the reader might gain some understanding of how color fields are related to baryon transport. By deriving expressions that mimic Gauss's Law, it was found that one can find quantities that behave like flux within the tube that induce baryon currents. Further, these fluxes are set by the quantum numbers of the color charge multiplet within the volume driving the flux. The $(p,q)$ values were found to play pivotal roles in determining both the tube's energy density and the baryon polarization or current.

The final sections apply the insight from the formal section to investigate the behavior of compound flux tubes, where the color charge multiplet to each side of some given point was defined by higher values of $p$ and $q$. Typically multiple partons must pass a given point for the point to return to a color singlet. It was show that if a highly excited tube is only allowed to decay via quark-antiquark production, and that if some point is characterized by $|p-q|\ge 3$, that baryonic charge would likely pass through the point in order for it to decay back to a color singlet. However, if the decay of the tubes were allowed to proceed through gluon transfer, then little baryon current might ensue from a highly excited tube. If the tube was created by a large number of quarks at one end, it was found that the typical generated values of $p-q$ did not grow noticeably with adding more quarks beyond three. Thus, baryon transport from the target and projectile region is unlikely to significantly increase once tubes have merged.

The last section considered the impact of adding gluonic charges along the flux tube. A model where $N_g$ random color charges were placed along the tube with the constraint that the overall state was a color singlet.  The energy density of the tube, which is characterized by $Q^{(2)}$, as a function of the position along the tube was found to behave similar to a random walk with $\approx N_g$ steps that is required to return to the origin. The energy density was found to be higher in the center of the tube, with the increase being mainly determined by $N_g$. I.e. for one or two gluons the energy density would be nearly constant, whereas as $N_g$ increases the central density increases while the energy density at the ends stays roughly fixed. If one end of the tube had more quarks, as one would expect in $pA$ collisions, the energy density was higher on the end with more quarks. Both the maximizing at mid-rapidity and the asymmetric rapidity multiplicity distributions in $pA$ are in qualitatively in line with experimental observation. In the Random Color Ansatz (RCA) the strength of both phenomena can be adjusted by varying $N_g$.

In addition to baryon transport from the target or projectile regions toward mid-rapidity, baryon transport plays a role in baryon-antibaryon correlations. The final topic of the last section concerned how the color fields in a flux tube can separate baryonic charge. In the context of the RCA, it was found that the separation between balancing baryons could be higher if the gluons were further separated, or if the flux tube was in a higher energy state, i.e. characterized by higher $p+q$ or $Q^{(2)}$. The ratio of separations is expected to fall off exponentially, but measuring the longer-range tail of the baryon-antibaryon balance function would require a large experimental acceptance.

It is difficult to estimate the degree to which the flux tube paradigm discussed here plays a role in baryon transport. Other mechanisms, such as the hard scattering of valence quarks might also play a part. Further, the degree to which one can consider multiple flux tubes as being distinct, vs. merged, is rather cartoonish. The approximation that tubes have a uniform field within a universal size, regardless of the excitation level, cannot be totally realistic. The RCA, which assumed the tube contains point-like color excitations from gluons, ignores the fact such gluons might be moving due to their transverse momentum. The idea that the initial gluonic configuration is created without regard to the strength of the local color fields, and that they then decay as if gluonic excitations are forbidden, is certainly an exaggeration.  Nonetheless, as illuminated in Sec. \ref{sec:kubo}, color fields exist once gluons are exchanged. Gluons can be radiated at intermediate rapidities and they should contribute to the strength of the color field. They provide energy density and forces which couple to baryonic charge. The energy density and baryonic forces should indeed be determined by the color state of the matter at that point. Thus, there is certainly at least some bit of truth to the phenomenology presented here, so the question of whether this phenomenology is true or false is naive. Instead, the question is in how much faith should one put in this phenomenology, and how much predictive power might such a picture provide. These questions requires a more detailed comparison with experiment, particularly with the measurement of baryons. 

Simple and compound flux tubes are not too difficult to picture in a $pp$ or in a $pA$ collision, but become much more difficult to define in a heavy ion collisions. In a heavy-ion environment, if one produces a quark-gluon plasma, the flux tubes would not be required to produce quark-antiquark pairs and transport the quarks to immediately form color singlets. Thus, it is hard to see how the lessons presented here could be relevant to heavy-ion collisions. Since most of the baryon transport from the target and projectile is expected to be related to the merging of color flux tubes, and if the tubes from a given baryon are like to merge closer to the baryon than the merging of tubes from different baryons, one might expect a similar amount of baryon transport in the initial stage for $pp$, $pA$ and $AA$ collisions. If there exists an exponential tail of the baryon transport, $dN_B/dy \approx Ae^{-\alpha|y-y_{\rm beam}|}$, subsequent diffusion would leave $\alpha$ unchanged while increasing the strength of the tail, $A$. However, the presumption that tubes from within the same baryon merge before merging with those of different baryons is rather speculative. 

Perhaps the best test of whether flux tubes are enhanced by intermediate gluons would be to study correlations in relative rapidity with a focus on larger relative rapidities, $\Delta y\gtrsim 1$. As is illustrated in Fig. \ref{fig:sampletrajectories}, the presence of a gluon is often accompanied by a change in the energy density of the tube. Thus, a detailed measurement of transverse energy correlations should be insightful. Two-point correlations binned by rapidity could involve transverse energy, multiplicity or baryon number. A baryon's rapidity could point to the position of an initial gluon, which would then point to a place at which the energy density was changing. At this point, the perspective proposed here is mostly speculative. However, a host of measurements, some of which are mentioned here, have the potential to evaluate the power of the color-flux-tube paradigm, particularly as to whether the locations of gluonic charges at intermediate rapidities drive the distributions and correlations of energy and baryonic charge.

\begin{acknowledgments}
This work was supported by the Department of Energy Office of Science through grant number DE-FG02-03ER41259. Conversations with Bj\"orn Schenke and Christian Bierlich are gratefully acknowledged.
\end{acknowledgments}

\bibliography{btransport}

\end{document}